\documentclass[a4paper]{article}

\usepackage[T1]{fontenc}
\usepackage{microtype}
\usepackage[utf8]{inputenc}

\usepackage{anysize}
\usepackage{amsmath}
\usepackage{amsfonts}
\usepackage[english]{babel}
\usepackage{url}
\usepackage{hyperref}
\usepackage{subcaption}
\usepackage{lmodern}

\usepackage{graphicx}
\usepackage{tabularx}
\usepackage{amssymb}
\usepackage{amsthm}
\usepackage{dsfont}

\usepackage{makecell}
\usepackage{booktabs} 
\usepackage{xcolor}
\usepackage[official]{eurosym}

\usepackage[ruled,vlined,noend]{algorithm2e} 

\newcommand{\breakingcomma}{%
  \begingroup\lccode`~=`,
  \lowercase{\endgroup\expandafter\def\expandafter~\expandafter{~\penalty0 }}}
\usepackage{breqn}

\title{Dynamic Traffic Assignment for Public Transport\\ with Vehicle Capacities} 

\author{Julian Patzner and Matthias Müller-Hannemann\\[1ex]
Institute of Computer Science\\
Martin Luther University Halle-Wittenberg, Germany\\
julian.patzner@student.uni-halle.de\\
matthias.mueller-hannemann@informatik.uni-halle.de
}
\date{}

\begin{document}

\thispagestyle{empty}
\maketitle

\begin{abstract}
Traffic assignment is a core component of many urban transport planning tools. It is used 
to determine how traffic is distributed over a transportation network. 
We study the task of computing traffic assignments for public transport: Given a public transit network, a timetable, vehicle capacities and a demand (i.e. a list of passengers, each with an associated origin, destination, and departure time), the goal is to predict the resulting passenger flow and the corresponding load of each vehicle. 
Microscopic stochastic simulation of individual passengers is a standard, but computationally expensive approach. Briem et al. (2017) have shown that a clever adaptation of the Connection Scan Algorithm (CSA) can lead to highly efficient traffic assignment algorithms, but ignores vehicle capacities, resulting in overcrowded vehicles. Taking their work as a starting point, we here propose a new and extended model that guarantees capacity-feasible assignments and incorporates dynamic network congestion effects such as crowded vehicles, denied boarding, and dwell time delays. Moreover, we also incorporate learning and adaptation of individual passengers based on their experience with the network.
Applications include studying the evolution of perceived travel times as a result of adaptation, the impact of an increase in capacity, or network effects due to changes in the timetable such as the addition or the removal of a service or a whole line. 
The proposed framework has been experimentally evaluated with public transport networks of Göttingen and Stuttgart (Germany). The simulation proves to be highly efficient. On a standard PC the computation of a traffic assignment takes just a few seconds per simulation day.\\[1ex]  

\end{abstract}

\section{Introduction}
Efficient, sustainable, and accessible public transport systems are critical to promoting economic growth, reducing congestion and minimizing environmental impact. This calls for innovative methods to optimize resource allocation, improve passenger comfort and ensure the overall efficiency of transit networks. A crucial part in the planning process of public transit systems is traffic assignment. Traffic assignment models are used to predict the passenger flow and the estimated load of vehicles within a transit network for a given demand scenario, making them a fundamental analysis and evaluation tool at both planning and operational levels~\cite{cats2020}. Results of traffic assignments provide valuable insights into possible congestion problems due to insufficient capacity. They can be used to study the benefits of introducing additional services, increased frequencies, larger vehicle capacities or possible network extensions~\cite{cats2016}.  
In this work, we consider the following variant of public traffic assignment: As input we are given a public transit network, a corresponding timetable and a vehicle schedule with vehicle capacities. The demand is specified by a list of passengers, each with an associated origin, destination, and departure time. 
The task is to assign for each individual passenger a journey from his origin to his destination. 

Microscopic stochastic simulation of individual passengers is meanwhile a standard, but computationally expensive approach. Briem et al.~\cite{briem2017} have shown that a clever adaptation of the Connection Scan Algorithm (CSA)~\cite{dibbelt2017} can lead to highly efficient traffic assignment algorithms. However, their approach ignores vehicle capacities, resulting in unrealistic assignments and overcrowded vehicles (they report in their case study assignments of about 1200 passengers to a single vehicle). 
The commercial state-of-the-art tool PTV VISUM has recently integrated CSA into their transport assignment for faster shortest path search~\cite{PTV2024}.

\paragraph*{Contribution.}
 Taking the work of Briem et al.~\cite{briem2017} as a starting point, we here propose a new and extended model that guarantees capacity-feasible assignments. We use agent-based modeling, a powerful tool to study the behavior of passengers, transport vehicles and the interaction between them. By modeling passengers as autonomous agents, this approach captures the different decision-making processes, preferences and adaptive behaviors that individuals exhibit during their journeys.
More specifically, our model incorporates dynamic network congestion effects such as crowded vehicles, denied boarding, and dwell time delays. Moreover, we also incorporate learning and adaptation of individual passengers based on their experience with the network. 
The proposed model has been implemented as a prototype. Computational experiments with public transport networks of Göttingen and Stuttgart (Germany) demonstrate the efficiency of the approach. 
 We present three case studies with selected applications:
\begin{enumerate}
	\item First, we study how passengers respond to network congestion. We find that the learning process is quite effective. It helps to improve the average perceived travel times and to reduce cases of denied boardings due to overcrowded vehicles.
	\item Second, we examine the benefits of increasing capacity. It turns out that a moderate increase in capacity leads to a significant reduction in average perceived travel times.
	\item Third, we compare unlimited vs.\ limited vehicle capacity. As expected the passenger flows with unlimited vehicle capacity turn out to be highly unrealistic. 
\end{enumerate}

\paragraph*{Related work.}
There is a long history of research on traffic assignment in public transport, see \cite{FU20121145,gentile2016,gentile2016b} for surveys. Conventional traffic assignment models distinguish between frequency-based and timetable-based models. These two groups differ in the modeling of the network. In frequency-based models~\cite{nguyen1988,OLIKER2020100005,spiess1989,schmoecker2011} the timetable is only modeled at the line level. Each line has an assigned frequency. These models aim at determining the average loads on the lines. In timetable-based models~\cite{hamdouch2008,hamdouch2014,nielsen2006,nuzzolo2012,nuzzolo2001}, the trips of a line are explicitly modeled, and the task is to determine loads on each single trip. A prominent example of the implementation of a schedule-based model is the commercial software VISUM, which is primarily used for long-term planning.
In agent-based models, passengers are not considered as an aggregated flow, but are simulated individually on a microscopic level. 
The individual vehicles are also modeled as individual agents, which allows great freedom in modeling (for example, the development of vehicle-specific delays or seating and standing capacities). Agent-based models focus on the dynamic interactions between passengers and the network as well as interactions between passengers. Individual, adaptive decisions are simulated as a reaction to dynamic network conditions. The network conditions are in turn dependent on the individual decisions of the passengers. 
In addition to dynamic processes within a day, a learning process lasting several days is usually modeled. The experiences on one day are incorporated into the expectations of the individual passengers and thus influence the decisions on subsequent days. These learning processes model long-term adaptations of passengers to the network conditions. 

In 2008, Wahba presented MILATRAS~\cite{wahba2008,wahba2014}, the first agent-based simulation in public transport that models a learning process. In MILATRAS, the traffic assignment is considered as a Markov decision problem, where the possible positions of the passengers (stops, vehicles) are the states and the possible decisions (choice of the next line or stop to alight) are the actions.  
In 2009, MATSim, an activity-based agent simulation framework, was extended by Rieser \textit{et~al.} to include public transportation trips~\cite{rieser2009}. In MATSim, each traveler has a population of plans representing journeys.  Each passenger randomly selects a plan from its population. This selection is based on journey ratings and the learning process is implemented as a co-evolutionary algorithm. 

With BusMezzo~\cite{cats2011,cats2020,cats2016} another agent-based simulation was introduced by Cats in 2011, designed as an operations-oriented model for short-term to mid-term planning~\cite{CatsHartl2016}. BusMezzo is implemented within the road traffic simulation Mezzo~\cite{burghout2006}. 
The probabilistic decisions in BusMezzo depend on the current expectations of passengers, based on previous days' experiences and current real-time information. The individual decisions (boarding, alighting and walking) depend on pre-computed path sets, where each action (e.g. alighting at a specific stop or boarding a specific vehicle) is assigned a path set (e.g. a subset of all possible paths to the destination after alighting). In~\cite{cats2020} a path is defined as a sequence of stops, whereby the exact lines and transfers are not specified. A single path is therefore not a concrete journey. The expected waiting time at a stop is calculated based on the combined frequency of the lines at the stop. 
A SoftMax model is used when deciding between different actions. Passengers learn the perceived travel times and the waiting times for the individual path segments. 
In contrast to MILATRAS and BusMezzo, the model proposed in this paper avoids the static pre-computation of alternative path sets. Instead, we consider and evaluate all feasible actions dynamically on-the-fly in an event-based manner, allowing passengers to react in a flexible way on network conditions such as unexpected delays or congestion. 
In our model the evaluation of individual passenger decisions depends on explicit journeys, which include specifically defined trips and transfers. Due to the explicit definition of journeys, the model is also suitable for timetables that include routes with low frequencies or individual special trips. 
Passengers can learn the expected load and reliability of specific trips, not only about lines. Similarly, probabilities of failed transfers can be learned.   

\paragraph*{Overview.} The remainder of this paper is structured as follows. First, we start with the necessary preliminaries to formalize the problem in Section~\ref{sec:basics}. In Section~\ref{sec:model}, we introduce our framework in detail.
 In Section~\ref{sec:experiments}, we present a computational study evaluating our framework and showcasing a few applications. Finally, we conclude with a short summary. 
 
%-------------------------------------------------------------------
\section{Preliminaries}
\label{sec:basics}

This section describes the modeling of the network and provides basic definitions and notations. A timetable is modeled as an event-activity network~\cite{hannemann2017}. 
An \emph{event-activity-network} is a tuple $(\mathcal{E}, \mathcal{A}, \mathcal{S}, \mathcal{T}, \mathcal{L},$ $\mathcal{F}, \mathcal{D})$ whose components are described below. The \emph{events} $\mathcal{E}$ and the \emph{activities}~$\mathcal{A}$ form a network $N = (\mathcal{E}, \mathcal{A})$, where the events correspond to the nodes and the activities to the arcs. The events are divided into \emph{departure events} $\mathcal{E}_{dep}$ and \emph{arrival events} $\mathcal{E}_{arr}$. Each event $e$ is associated with a time $\tau(e)$, a \emph{trip} $trip(e)\in\mathcal{T}$, and a \emph{stop} $stop(e)\in\mathcal{S}$. We write $dep(s)$ and $arr(s)$ for the set of all departure and arrival events at stop $s$.
An \emph{activity} $(e_1, e_2)$ can be a driving, dwelling or transfer activity. 
 A \emph{dwelling arc} is an arc from an arrival event to a departure event, modeling the waiting of a vehicle at a stop. \emph{Driving arcs} are arcs from a departure  to an arrival event and model driving from one stop to the next. Note that driving arcs are called \emph{(elementary) connections} in the context of the Connection Scan Algorithm.
  Driving and dwelling activities have an \emph{in-vehicle time} $\tau^{ivt}(e_1, e_2) := \tau(e_2)-\tau(e_1)$ and a \emph{minimum in-vehicle time} $\tau^{ivt}_{min}(e_1, e_2)$. The difference between the regular duration and the minimum duration of an arc corresponds to the catch-up potential in case of delays. A driving or dwelling activity $(e_1, e_2)$ has a reference to its trip $trip(e_1, e_2)\in\mathcal{T}$.

A \emph{trip} $t\in\mathcal{T}$ is an alternating sequence of departure and arrival events $\breakingcomma (e_{dep}^1(t), e_{arr}^2(t), e_{dep}^2(t),. ..., e_{arr}^{|S(t)|-1}(t), e_{dep}^{|S(t)|-1}(t), e_{arr}^{|S(t)|}(t))$, where $S(t)$ is the set of stops served by the trip~$t$. Denote by $e_{dep}^i(t)$ the departure event and by 
$e_{arr}^i(t)$ the arrival event at the $i$th stop of $t$. The times of the events of a trip are non-decreasing, so $\tau(e_{dep}^i(t))\leq \tau(e_{arr}^{i+1}(t))$ and $\tau(e_{arr}^i(t))\leq \tau(e_{dep}^i(t))$ always apply. This sequence of events defines the alternating sequence of driving and dwelling $activities(t)$ of trip $t$. For a \emph{trip segment} between the $i$th and $j$th stop of a trip $t$, with $i<j$, we write $e_{dep}^i(t)\rightarrow e_{arr}^j(t)$. This trip segment contains all driving and dwelling arcs between the departure event at the $i$th stop and the arrival event at the $j$th stop of trip $t$. Let $activities(e_{dep}^i(t)\rightarrow e_{arr}^j(t))$ be this sequence of arcs. 
Each trip has a \emph{seat capacity} $cap_{sit}(t)$, which corresponds to the number of seats in the vehicle. The \emph{capacity} $cap(t)\geq cap_{sit}(t)$ of a trip is the sum of the seats and standing capacity. This capacity is considered as a hard upper limit for the number of passengers that can be on a trip. Trips are grouped into \emph{lines} $\mathcal{L}$, where each trip of a line serves the same sequence of stops. Let $line(t)$ be the line of a trip $t$ and $line(e)$ the line of an event $e$. We assume that two trips of a line cannot overtake each other. A line is therefore a set of trips ordered according to the first departure time. Let $t_1$ and $t_2$ be two subsequent trips of a line. The \emph{headway} $headway(e_{x}^i(t_1)):=\tau(e_{x}^i(t_2))-\tau(e_{x}^i(t_1))$ of an event is the time until the corresponding event of the next trip.
A \emph{transfer} is an arc from an arrival event to a departure event of another trip. Such arcs are not explicitly modeled, but are implicitly defined by the stops $\mathcal{S}$ and \emph{footpaths}~$\mathcal{F}$. A footpath $(s, s')\in \mathcal{F}$ between two stops can be passed at any time. The time required for a footpath is given by $\ell(s, s')\in \mathbb{N}$. A \emph{minimum transfer time} $mct(s)$ can be specified for transferring at stop $s$. 
A transfer $e_{arr}\rightarrow e_{dep}$ with $trip(e_{arr})\neq trip(e_{dep})$ is therefore \emph{valid} if either $stop(e_{arr}) = stop(e_{dep})$ and $\tau(e_{arr}) + mct(stop(e_{arr})) \leq\tau(e_{dep})$ applies, or if $stop(e_{arr})\neq stop(e_{dep})$ and the footpath $(stop(e_{arr}), stop(e_{dep}))$ exists with $\tau(e_{arr}) + \ell(stop(e_{arr}), stop(e_{dep})) \leq\tau(e_{dep})$.
A valid transfer can become invalid in the course of the simulation due to delays. Conversely, an invalid transfer can also become valid if the departure event is delayed. The \emph{walking time} $\tau^{walk}(e_{arr}\rightarrow e_{dep})$ of a transfer is $\ell(stop(e_{arr}), stop(e_{dep}))$ if $stop(e_{arr})\neq stop(e_{dep})$, and $0$ otherwise. The \emph{waiting time} $\tau^{wait}(e_{arr}\rightarrow e_{dep})$ of a transfer is $\tau(e_{dep}) - \tau(e_{arr}) - \tau^{walk}(e_{arr}\rightarrow e_{dep})$.
To model delays that can propagate between different trips, \emph{dependency arcs} $(t_1, t_2)\in\mathcal{D}$ are introduced between two consecutive trips of a vehicle. Like activities, they have a minimum duration. If a trip arrives late at its last stop, the next trip of the vehicle is delayed accordingly. 

The agents are generated using an OD-matrix. The OD-matrix specifies how many passengers per hour want to travel from a specific start stop $origin$ to a specific destination stop $dest$. Each passenger is assigned a fixed start time $\tau_{start}$. 
A more realistic modeling, in which the start time is chosen by the agents themselves depending on the network, is conceivable, but is not dealt with in this paper. 
During a simulated day, a \emph{journey} is created for each passenger, which is an alternating sequence of trip segments and valid transfers. In addition to transfers between two trips, a journey can also have an initial walk at the start or a final walk to the destination. We therefore extend our definition of transfers to include the special cases $origin\rightarrow e_{dep}$ and $e_{arr}\rightarrow dest$, where $origin$ is the start and $dest$ is the destination. A final transfer $e_{arr}\rightarrow dest$ always has a waiting time of~$0$. A journey $J$ consisting of $n$ trip segments therefore has the form $\breakingcomma J=\{origin\rightarrow e_{dep}^{i_1}(t_1), e_{dep}^{i_1}(t_1)\rightarrow e_{arr}^{j_1}(t_1),e_{arr}^{j_1}(t_1)\rightarrow e_{dep}^{i_2}(t_2),. ..., e_{dep}^{i_n}(t_n)\rightarrow e_{arr}^{j_n}(t_n),e_{arr}^{j_n}(t_n)\rightarrow dest\}$.

%-------------------------------------------------------
\section{Agent-Based Dynamic Traffic Assignment Model}
\label{sec:model}

In this section we introduce our dynamic traffic assignment model step-by-step. We first sketch and discuss the model of Briem et~al.~\cite{briem2017}, as it serves in many respects as the basis for the simulation presented in this work. Then we present a high-level description of our simulation model. Afterwards, we provide details about the modeling of congestion effects, passenger preferences, route choice, learning, and real-time reactions.

\paragraph*{Traffic assignment using Connection Scan Algorithm.}

In~\cite{briem2017}, passenger preferences are modeled using perceived arrival times. These are used to make decisions based on factors such as arrival time, number of transfers, walking time, waiting time, and delay robustness. The algorithm simulates different decisions for each passenger (boarding a vehicle, alighting a vehicle, walking to another stop) and assigns probabilities to these options based on the perceived arrival times of each option. The boarding and alighting decisions are binary (board or stay at a  stop, alight or stay on a trip). The perceived travel times for all options are calculated with a single run of the Connection Scan Algorithm~\cite{dibbelt2017}. In a second scan over all elementary connections, a random decision whether to board the vehicle is made for each passenger waiting at the corresponding departure stop. Then, for all passengers in the current vehicle, a random decision is made whether to alight at the arrival stop. Finally, for each alighting passenger, a random decision is made as to which stop they will walk to (or remain at the current stop).

This approach is very efficient, but is based on some unrealistic assumptions. First, the model assumes unlimited vehicle capacities. This leads to traffic allocations in which individual vehicles have unrealistically high load factors. Second, passengers do not react in any way to high occupancy rates and are treated as uniform decision makers; personal preferences or experiences are not implemented. Third, movements of vehicles are not simulated and transfer uncertainty is modeled by adding a random variable to the arrival times. Consequently, it is not possible to model specific vehicle delays and how passengers react to them. As capacities and occupancy rates are ignored and delays are modeled as random variables, the network performance is completely independent from passenger decisions. The binary modeling of boarding and alighting decisions also leads to unrealistic traffic assignments, as decisions are biased towards earlier boardings and alightings. In a sequence of possible boardings, a decision for later journeys only takes place if the passenger has decided against boarding every previous journey. In this way, examples can be constructed in which the optimal journey has an arbitrarily small probability.

\subsection{High-level Description of Model}

In our model, passengers are modeled as agents with their own preferences and experiences. The individual decisions of the passengers influence the network dynamics and the passenger decisions are in turn dependent on the network dynamics. The network flow is therefore the result of the passengers' interaction with the network. Each individual trip is modeled as a separate entity whose performance depends on the decisions of the agents. The modeling of vehicles as entities allows explicit delays and delay dependencies between vehicles. Section~\ref{sec:Congestion Effects} describes three different network congestion effects that are implemented in this model: crowded vehicles (including seat allocations), denied boardings and dwell time delays. 

The model replicates the impact of network performance on passenger decisions. We develop a flexible choice model that allows passengers to make adaptive decisions in response to these dynamic network conditions. As in \cite{briem2017}, we evaluate decisions by calculating a perceived travel time (Section~\ref{sec:Perceived Travel Time}) for each option, but we incorporate the three modeled congestion effects. In Section~\ref{sec:choicemodel}, we explain how probabilities are assigned to different options based on the perceived travel times. Instead of binary choices, passengers choose between boarding trips of different lines and alighting at different downstream stops. The advantage of this is that passengers can react adaptively to the characteristics of different journeys, rather than just considering the current and optimal options. One drawback is that the journey characteristics can change in the time between the decision and the execution of the chosen event. In Section~\ref{sec:Real-time reactions}, we describe how passengers can change their decisions based on real-time information. 

Individual passengers adapt their behavior to their experiences with the network. Passengers' decisions are influenced by the experience they have gained by repeatedly traveling through the network on consecutive days (Section~\ref{sec:Learning}). Modeling a learning process is of great importance as the network conditions vary from day to day. The learning process models how passengers react to the fluctuating network congestions on different days and allows them to avoid highly congested trips.

\begin{algorithm}[tb]
\DontPrintSemicolon
\caption{main loop of the simulation}\label{alg:mainloop}
calculate initial perceived travel times $f_k$\;
$day\gets 0$\;
\While{$day < number\_of\_days$}
{
	$Q\gets getEventsWithinSimulationFrame()$\;
	\tcc{Events in $Q$ are sorted by current time and event type as a secondary criterion (arrival before departure events)}
	\While{$Q\neq\emptyset$}
	{
		$event\gets Q.top()$; $Q.pop()$\;
		$current\_time\gets \tau(event)$\;
		\For{unstarted passenger $k$ with $\tau_{start}(k)\leq \tau(event)$}
		{
			$k.makeInitialDecision()$\;
		}
		\If{$event$ is departure event}
		{
			\ForAll{passengers $k$ at $stop(event$) in random order}
			{
				$k.executeDepartureEvent()$\;
			}
			$createAndPropagateDwellingDelay(event)$\;
			$Q.update()$\;
		}
		\If{$event$ is arrival event}
		{
			\ForAll{passengers $k$ in $trip(event)$}
			{
				$k.executeArrivalEvent()$\;
			}
		}
	}
	\ForAll{passengers $k$}
	{
		update expectations $\tilde{\lambda}_k, \tilde{p}_k^{denied}$ and $\tilde{\tau}_k$\;
		update perceived travel times $f_k$
	}
	$day\gets day+1$\;
}
\end{algorithm}

\begin{figure}[tb]
\centering
\includegraphics[scale=0.4]{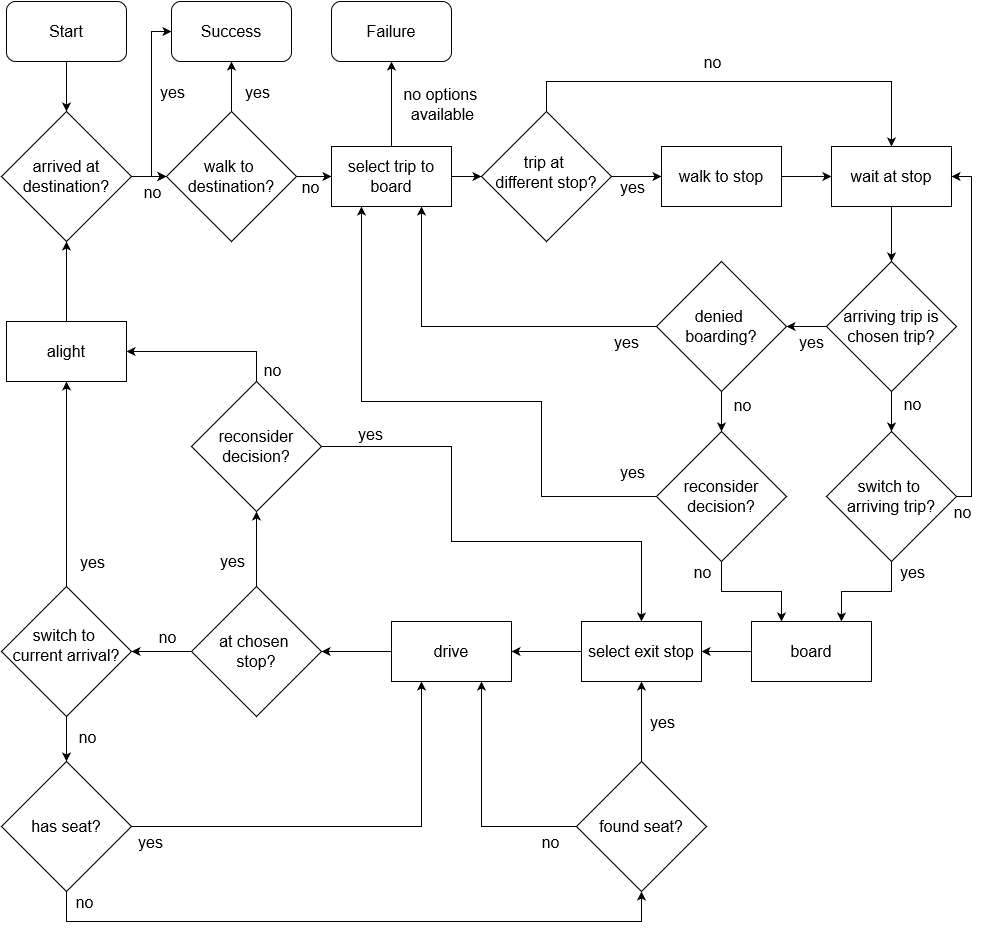}
\caption{Flow chart from the passengers' perspective for one day.}
\label{fig:flowchart}
\end{figure}

The model is developed as an event-oriented simulation with discrete time steps. We keep the algorithmic framework and the overall structure of a simulated day from the model in \cite{briem2017}. Changes to the algorithm are described in Section~\ref{sec:Perceived Travel Time}. Algorithm~\ref{alg:mainloop} shows the main loop of the simulation. 
   As event times are no longer static (in particular, due to dwell time delays), the connections cannot be pre-sorted and we need a priority queue~$Q$. This contains all events of the period to be simulated. The events are sorted in ascending order according to the current time of the events. If two events have the same time, arrivals are processed before departures. 
On each pass through the main loop, the current event is extracted from~$Q$. All passengers who have not yet started their day are loaded into the network. They select their first boarding. 
The type of event is then distinguished. 
In the case of a departure event, the passengers waiting at the current stop are processed in random order. In the case of an arrival event, the passengers on the current vehicle are processed. The processes for the passengers are described in more detail in Figure~\ref{fig:flowchart}. 
When a departure event is processed, a dwell time delay may be generated and propagated. In this case, $Q$ must be updated. At the end of each day, the expectations and perceived travel times are updated.

\subsection{Congestion Effects}
\label{sec:Congestion Effects}

\paragraph*{Crowded vehicles.} Seats are allocated as follows: We assume that passengers alight from the vehicle before those waiting at the stop board. Seats may therefore become available. First, passengers are drawn at random from those currently standing until either all seats are occupied or all passengers are seated. Second, the waiting passengers then board the vehicle in random order. We therefore assume that the passengers mix while waiting. Entering passengers are assumed to take a seat if one is available.
The main causes of dissatisfaction in overcrowded vehicles are standing and physical proximity to other passengers. Let $q^{onboard}(e_1, e_2)\leq cap(t)$ be the number of passengers of an activity $(e_1, e_2)$ of the trip $t$. The discomfort depends on the current passenger \emph{load} $\lambda(e_1, e_2):=q^{onboard}(e_1, e_2)/cap_{sit}(t)$ and on whether a seat has been found. The load is defined relative to the number of seats. These two properties give the \emph{crowding factor} $\beta_k^{crowding}(\lambda(e_1, e_2),seated_k(e_{dep},(e_1, e_2)))$, where $k$ is the current passenger and $e_{dep}$ is the departure event at which the passenger boarded the current trip. The Boolean function $seated_k$, which indicates whether the passenger $k$ has a seat, thus depends not only on the current arc, but also on the time of boarding. 
We assume that a passenger will not give up a seat once it has been found.
 The crowding factor is modeled as a step function (see Table~\ref{tbl:crowdingpenalty}). The load of a vehicle is not known to the passenger in advance and is therefore based on the passenger's personal experience on previous days, or a default value if no experience is available. Learning is described in Section~\ref{sec:Learning}. The model also allows for real-time load information, but this is beyond the scope of this paper.

\paragraph*{Denied boardings.} As with the seat allocation, the standing room allocation depends on the random order of boarding passengers. It is possible that the number of passengers wishing to board a vehicle is greater than the remaining capacity. In this case, some of the passengers must therefore remain at the stop. We refer to this as \emph{denied boarding} at a departure event $e_{dep}$. The passengers must respect the minimum transfer time at the current stop and may walk to another stop in response to the denied boarding. They are therefore treated as passengers who alight at $stop(e_{dep})$ at time $\tau(e_{dep})$. 
Such an unplanned complication is associated with additional stress for the passenger. 
Therefore, the subsequent waiting or walking time to the next boarding event is penalized by a multiplier $\beta_k^{fail}$.

\paragraph*{Dwell time delays.} We model the dwell time as a monotonically increasing function based on the number of boarding and alighting passengers. Let $q^{alight}$ be the current number of alighting passengers and $q^{board}$ the current number of boarding passengers. The required dwell time is given by 
$(q^{alight} + q^{board})/{doorCapacity(t)}$,
where $doorCapacity(t)$ is the number of passengers that can board or alight per second. This value differs significantly between different vehicle types, for example buses generally have a smaller $doorCapacity$ than trains. It should be noted that this is a greatly simplified model; for example, the time to open and close the doors is ignored. Since boarding and alighting is the dominant component~\cite{nap}, this is sufficient to capture the systematic evolution of delays caused by the network flow. For more accurate modeling, more information is needed on the vehicles used.
If the required dwell time is greater than the scheduled one of a dwelling arc, the corresponding departure is delayed by the difference. 
Occurring delays are propagated downstream along the corresponding trip (as in \cite{SchieweSchoebelJaegeretal.2023}). Additionally, delays of trains are propagated across shared rails.

\subsection{Perceived Travel Time}
\label{sec:Perceived Travel Time}

Perceived travel time is a key characteristic that influences passenger satisfaction with public transport. Unlike actual travel time, it takes into account that waiting times, walking distances, transfers and in-vehicle crowding are perceived differently by passengers.
Each passenger in the model has its own preferences and experiences. The perceived travel time is dependent on the network dynamics of the current day and the passenger's experience gained on previous days. Each passenger has different sources of information about expected times and vehicle loads. These sources can be, in descending order of priority, real-time information, experience, or default values (scheduled times or an input parameter for load). Unless otherwise specified, the source with the highest priority is used. 
 
  During the simulation, each passenger is assigned a journey iteratively through partial decisions. 
Two types of decisions are made: a passenger waiting at a stop, has to decide which trip to board next (or whether to walk directly to the destination if there is a footpath). 
The passenger chooses between departure events that can be reached from the current stop, including departure events reachable by a footpath.
Second, a passenger traveling in a vehicle has to decide at which stop to alight. 
To implement these decisions, we calculate for each passenger $k$ an expected \emph{perceived travel time} $f_k(e)$ for each boarding and alighting event $e$, that corresponds to the optimal journey from the current event to the destination. The initial perceived travel times are calculated before the start of the simulation and are based on scheduled times. The perceived travel times are updated at the end of each day after incorporating passenger experience, and are recalculated during a day when real-time information is available. At the time of the decision, the passenger does not know the actual perceived travel times, as the activities are in the future. These values are therefore only estimates for the current day. 

We account for crowding by weighting the in-vehicle-travel time by a crowding factor $\beta_k^{crowding}$ depending on the vehicle load. Similarly, waiting and walking times are weighted by passenger-specific factors $\beta_k^{walk}$ and $\beta_k^{wait}$, respectively. We also use an additive penalty $\beta_k^{transfer}$ for each transfer and $\hat{\beta}_k^{fail}(tr)$ for a possible failed transfer~$tr$. The penalty term for a failed transfer corresponds to the weighted additional waiting time caused by the failed transfer, multiplied by an estimated probability $p_k^{fail}(tr)$. The expected perceived travel time $f_k(e)$ of an event $e$ is defined recursively as the minimum over all possibilities to continue from this event to the destination. In the following, we derive these calculations step by step.

First, we define an expected perceived travel time $ptt_k$ for each transfer and for each trip segment. The perceived travel time of a journey is the sum of the perceived travel times of all trip segments and transfers of the journey (including the waiting time at the origin stop). The perceived travel time of a trip segment is the sum of the perceived travel times of all driving and dwelling arcs of the trip segment. The perceived travel time of an activity $(e_1, e_2)$ is obtained by multiplying the crowding factor with the duration of the activity, i.e.
\[ ptt_k(e_{dep},(e_1, e_2)) := \beta_k^{crowding}(\lambda_k(e_1, e_2),seated_k(e_{dep},(e_1, e_2)))\cdot \tau_k^{ivt}(e_1, e_2) ,\]
where $e_{dep}$ is the departure event at which the passenger $k$ boarded the current trip. Since a passenger does not know in advance when he will find a seat, he assumes that he will find a seat at the first arc with an expected load of less than $1$. The Boolean function $seated_k(e_{dep}^i(t),(e^j(t),e^{j'}(t)))$ is $true$ if an arc $(e^m(t),e^{m'}(t))$ exists with $\lambda_k(e^m(t),e^{m'}(t))<1$ and $i\leq m\leq j$. This is a pessimistic estimate: even if the expected loads match the actually experienced values, the passenger may find a seat earlier.

The perceived travel time of a trip segment is
\[ ptt_k(e_{dep}^i(t)\rightarrow e_{arr}^j(t)) := \sum_{(e, e')\in activities(e_{dep}^i(t)\rightarrow e_{arr}^j(t))} ptt_k(e_{dep}^i(t),(e, e')).\]

The real travel time of a transfer $tr=e_{arr}\rightarrow e_{dep}$ consists of the waiting time for the next trip and, if a footpath is required for the transfer, the length of the footpath. These times are multiplied by passenger-specific coefficients $\beta_k^{walk}$ and $\beta_k^{wait}$. In addition, there are penalty terms $\beta_k^{transfer}$ for the transfer itself and $\hat{\beta}_k^{fail}(tr)$ for a possible failed transfer. The penalty term for a failed transfer corresponds to the weighted additional waiting time caused by the failed transfer. Since the additional waiting time after the failed transfer is not known at the time of the decision, it must be estimated. To allow an efficient calculation, this estimate is based only on the timetable. We define the expected additional weighted waiting time after the failed boarding $\hat{\beta}_k^{fail}(tr)$ as $headway(e_{dep})\cdot \beta_k^{fail}$. The value $\hat{\beta}_k^{fail}(tr)$ is then multiplied by a probability $p_k^{fail}(tr)$ estimated by passenger $k$ that boarding at $e_{dep}$ is not possible due to limited capacity of $trip(e_{dep})$ or delays of $e_{arr}$. This probability is assumed to be $0$ at the beginning of the simulation. It depends on two components, $p_k^{denied}$ and $p_k^{delay}$. The overall probability $p_k^{fail}(tr)$ is calculated as $p_k^{denied}(e_{dep}) + p_k^{delay}(tr) - (p_k^{denied}(e_{dep}) \cdot p_k^{delay}(tr))$. Both components are based on the passenger's experience on previous days. The probability $p_k^{delay}$ is calculated using a weighted empirical distribution function of the arrival times $\tau(e_{arr})$. 

The perceived travel time of a transfer $tr=e_{arr}\rightarrow e_{dep}$ is
\[ptt_k(tr) :=\hat{\beta}_k^{fail}(tr)\cdot p_k^{fail}(tr) + \beta_k^{wait}\cdot \tau_k^{wait}(tr) +\beta_k^{walk}\cdot \tau^{walk}(tr) + \beta_k^{transfer}.\]
We additionally define $walk_k^{dest}(s) =\beta_k^{walk}\cdot\ell(s, dest)$ as the weighted walking time from stop $s$ to destination $dest$ of passenger $k$. Using these definitions, we can now recursively define the expected perceived travel times $f_k(e)$. The minimum perceived travel time from an arrival event $e_{arr}$ to the destination stop is given by the minimum over all outgoing valid transfers and the weighted walking time to the destination stop. Transfers that are only possible due to a learned delay of the departure event are ignored here. We first define the minimum over all transfers:
\[f^{trans}_k(e_{arr}) := \min_{e_{arr}\rightarrow e'_{dep}} ptt_k(e_{arr}\rightarrow e'_{dep}) + f_k(e'_{dep}).\]
The value $f_k(e_{arr})$ is the minimum of this value and the weighted walking time, i.e.
\[f_k(e_{arr}) = \min\{f^{trans}_k(e_{arr}), walk_k^{dest}(stop(e_{arr}))\}.\]
In particular, $f_k(e_{arr})$ is $0$ if $stop(e_{arr})$ is the destination of $k$. The minimum perceived travel time of a departure event $e_{dep}$ is the minimum over all possible trip segments. The respective trip segment is defined by the arrival event at which the passenger alights. For a departure event $e_{dep}^i(t)$ this results in 
\[f_k(e_{dep}^i(t)) = \min_{j>i} ptt_k(e_{dep}^i(t)\rightarrow e_{arr}^j(t))+f_k(e_{arr}^j(t)).\]
We have therefore defined a minimum perceived travel time $f_k(e)$ to the destination for each passenger $k$ and for each possible boarding and alighting event.

\paragraph{Calculation of initial perceived travel times.}

Algorithm~\ref{alg:init_profile} describes the calculation of the initial perceived travel times $f_k(e)$. These initial values are independent of the network dynamics. The times therefore correspond to the regular times according to the timetable, a standard load $\lambda_{std}$ is assumed for each activity and the probability of a failed boarding is assumed to be $0$. The algorithm is based on the profile connection scan algorithm~\cite{dibbelt2017}, with the difference that we calculate perceived travel times instead of earliest arrival times. In addition, dwelling activities must also be taken into account.
As in \cite{dibbelt2017}, we first perform a simple earliest arrival time query with CSA to determine the driving arcs $C$ that can be reached from the origin. We limit the time horizon to $\tau_{arr}(k)+\Delta_{\tau}$, where $\tau_{arr}(k)$ is the earliest (real) arrival time of $k$ at its destination $dest$. We therefore discard journeys that arrive more than $\Delta_{\tau}$ later at the destination than the fastest journey. 
During the execution of the algorithm, a set of Pareto-optimal journeys $B[s]$ from $s$ to the destination is calculated for each stop $s\in\mathcal{S}$. The criteria are the departure time and the minimum perceived travel time to the destination. For each departure event $e_{dep}$, a label $L=(\tau_{dep}, ptt, trip)$ is created consisting of the departure time $\tau_{dep}$, the minimum perceived travel time to the destination $ptt$ and the first trip of the journey $trip$. For Pareto dominance, the difference between the departure times of the labels must be added to the perceived travel time of the later label. 

We iterate over the driving arcs $C$ in descending order of departure times. In each loop iteration, the invariant applies that $ptt_{curr}[t]$ is the minimum perceived travel time from the earliest scanned departure event of the trip $t$ to the destination. At the beginning of the iteration for the travel arc $(e_{dep}^i(t), e_{arr}^{i+1}(t))$, $ptt_{curr}[t]$ therefore corresponds to the minimum perceived travel time from $e_{dep}^{i+1}(t)$ to the destination. At the arrival event $e_{arr}^{i+1}(t)$, the passenger has three options: he can change to another trip, walk to the destination or stay in the vehicle. We calculate the minimum of these three options. First, we calculate the perceived travel time for a transfer. The optimal transfer is the transfer to the Pareto-optimal partial journey with the smallest departure time. Here we discard the case that a passenger alights to get back on the current trip immediately. Let $L_f$ be the label of this journey. The perceived travel time for a transfer $transfer$ is therefore the sum of the perceived travel time of $L_f$ and the cost of the transfer (waiting time and penalty for transfer). The perceived travel time for the passenger to walk to the destination is given by $walk_k^{dest}(stop(e_{arr}^{i+1}(t)))$. We summarize these two options under $alight$. The perceived travel time for staying in the vehicle is equal to the sum of the minimum perceived travel time from $e_{dep}^{i+1}(t)$ to the destination (i.e. $ptt_{curr}[t]$) and the cost of the dwelling arc between $e_{arr}^{i+1}(t)$ and $e_{dep}^{i+1}(t)$. The sum of these two costs is called $remain$. The minimum of all three options is $minptt$. 

Afterwards, $ptt_{curr}[t]$ is updated. The minimum perceived travel time for the departure event $e_{dep}^i(t)$ corresponds to the sum of the costs of the current driving arc and the costs of the minimum option at the arrival event ($minptt$). We store the calculated minimum perceived travel times in $f_k$. 
We still need to update the Pareto sets. We create a label for the current stop and for all footpaths. We incorporate the minimum transfer time or footpath length directly into the labels. The departure time of the label therefore corresponds to the departure time of $e_{dep}^i(t)$ minus the minimum transfer time or walking distance. For the perceived travel time $ptt$ of the labels, the corresponding costs for waiting or walking must be added to the current perceived travel time $ptt_{curr}[t]$. We first test whether a label is dominated by another label of the corresponding Pareto set. If this is not the case, it is inserted. The labels dominated by the inserted label are then removed.

\begin{algorithm}[tb]
\DontPrintSemicolon
\caption{Calculation of the initial perceived travel times $f_k$ to the destination of passenger $k$}\label{alg:init_profile}
\KwIn{list $C$ of relevant driving arcs sorted by regular departure times}
\KwOut{perceived travel times $f_k(e)$}
\ForEach{$t\in \mathcal{T}$}
{
	$ptt_{curr}[t]\gets \infty$\;
}
\ForEach{driving arc $(e_{dep}^i(t), e_{arr}^{i+1}(t))\in C$, descending by $\tau(e_{dep}^i(t))$}
{
	let $L_{transfer}$ be the label $L\in B[stop(e_{arr}^{i+1}(t))]$ with minimum departure time $\tau_{dep}(L)$, for which $\tau_{dep}(L)\geq \tau(e_{arr}^{i+1}(t))$ and $trip(L)\neq t$ apply\;
	\eIf{$L_{transfer}\neq \bot$}
	{
		$transfer\gets ptt(L_{transfer}) + \beta_k^{transfer} + \beta_k^{wait}(\tau_{dep}(L_{transfer})-\tau(e_{arr}^{i+1}(t)))$\;
	}
	{
	$transfer\gets\infty$\;
	}
	$alight\gets\min\{transfer, walk_k^{dest}(stop(e_{arr}^{i+1}(t)))\}$\;
	$remain\gets ptt_{curr}[t]+\beta_k^{crowding}(\lambda_{std}, \lambda_{std}<1)\cdot\tau^{ivt}(e_{arr}^{i+1}(t), e_{dep}^{i+1}(t))$\;
	$minptt\gets\min\{alight, remain\}$\;
	\lIf{$minptt=\infty$}
	{
		\textbf{continue}
	}
	\;
	$ptt_{curr}[t]\gets minptt + \beta_k^{crowding}(\lambda_{std}, \lambda_{std}<1)\cdot\tau^{ivt}(e_{dep}^{i}(t), e_{arr}^{i+1}(t))$\;
	$f_k(e_{arr}^{i+1}(t))\gets alight$\;
	$f_k(e_{dep}^{i}(t))\gets ptt_{curr}[t]$\;
	$L_s\gets(\tau(e_{dep}^i(t)) - mct(stop(e_{dep}^i(t))), ptt_{curr}[t] + \beta_k^{wait}\cdot mct(stop(e_{dep}^i(t))), t)$\;
	if $L_s$ is not dominated: insert $L_s$ into $B[stop(e_{dep}^{i}(t))]$ and remove dominated labels\;
	\ForEach{footpath $(s', stop(e_{dep}^{i}(t))\in \mathcal{F}$}
	{
		$L_f\gets(\tau(e_{dep}^i(t)) - \ell(s', stop(e_{dep}^{i}(t)), ptt_{curr}[t] + \beta_k^{walk}\cdot \ell(s', stop(e_{dep}^{i}(t)), t)$\;
		if $L_f$ is not dominated: insert $L_f$ into $B[s']$ and remove dominated labels\;
	}
}
\end{algorithm}

\paragraph{Updating the perceived travel times.}
Updating the perceived travel times $f_k$ works in a similar way to the initial calculation in Algorithm~\ref{alg:init_profile}. This subsection only describes the differences.
Only a small subset of all events is affected by the update. It would therefore be suboptimal to scan the entire list $C$. Instead, we use a priority queue $Q$, which contains the arcs in descending order according to regular departure times. At the beginning, $Q$ contains the arcs for which at least one property has changed. At the end of each iteration, all driving arcs through which the current driving arc can be reached are inserted into $Q$. These are the previous driving arc of the current trip $t$ and all driving arcs $(e_{dep}^{j}(t'), e_{arr}^{j+1}(t'))$ for which the transfer $e_{arr}^{j+1}(t')\rightarrow e_{dep}^i(t)$ is valid.
Another difference is that during the update, each stop is usually visited much less frequently, as only a small subset of the travel arcs are scanned. We therefore do not calculate the Pareto sets. Instead, we calculate the minimum perceived travel time for a transfer by scanning over all departure events $e_{dep}'$ that are reachable from $e_{arr}^{i+1}(t)$ via a valid transfer. Transfers that are only valid because of the learned delay of the boarding event are ignored. We therefore use the regular time for $e_{dep}'$. 

In the initial calculation, a default value $\lambda_{std}$ was assumed for the load of each arc. This means that a passenger assumes that they are either always seated or always standing. This is no longer the case with the update. As a reminder: A passenger assumes that they must stand until they reach an arc $(e_{dep}^i(t), e_{arr}^{i+1}(t))$ for which they expect a load factor of less than $1$, i.e. $\tilde{\lambda}_k(e_{dep}^i(t), e_{arr}^{i+1}(t))<1$. In addition to $ptt_{curr}[t]$, we store another value $ptt_{curr}^{sitting}[t]$, which is the minimum perceived travel time to the destination, assuming that the passenger is seated for the rest of the journey. If we scan an arc $(e_{dep}^i(t), e_{arr}^{i+1}(t))$ with $\tilde{\lambda}_k(e_{dep}^i(t), e_{arr}^{i+1}(t))<1$, we set $ptt_{curr}[t]\gets ptt_{curr}^{sitting}[t]$. 

The initial calculation of the minimum perceived travel times and their update at the end of the day or in real-time reactions is independent of the other passengers. The parallelization of these steps is therefore trivial.

\subsection{Choice Model}
\label{sec:choicemodel}
We use a mixed $(\epsilon-greedy, SoftMax)$ decision model. With a probability of $1-\epsilon$, the optimal decision is made directly. In the other case, the SoftMax selection is used. This mixed model results in the agents predominantly making the optimal decision, while occasionally opting for a random action according to the SoftMax principle. The SoftMax function is used to assign probabilities to the individual decisions based on the perceived travel times $f_k$.

In general, the SoftMax selection for any actions $a$ with costs $f(a)$ has the following form:
\[p(a):=\frac{e^{(f(a^{opt}) - f(a))/\gamma(d)}}{\sum\limits_{a'} e^{(f(a^{opt}) - f(a'))/\gamma(d)}},\]
where $a^{opt}$ is the optimal action, $\gamma(d)$ is the temperature and $d$ is the current day. The costs of an action are therefore considered relative to the optimal costs. If a high temperature is chosen, the probability of making suboptimal decisions is higher. In the limit value for $\gamma\rightarrow 0$, the optimal decision is always made. The temperature therefore influences the average perceived travel times of passengers.

\paragraph*{Boarding and walking decisions.}
When a passenger $k$ is waiting at a stop, he decides on a trip, specifically a departure event, which he wants to board next. 
For simplicity, assume that the 
passenger just alighted at an arrival event $e_{arr}$. 
The passenger decides on the basis of the perceived travel time $f_k$. We restrict the departure events in question to the earliest available trips on each line. Let the set of relevant departure events be $reldep(e_{arr}): = \{e_{dep}^i(t)|e_{arr}\rightarrow e_{dep}^i(t)$ is valid and there is no valid transfer $e_{arr}\rightarrow e_{dep}^i(t')$ with $line(t) = line(t')$ and $\tau(e_{dep}^i(t'))<\tau(e_{dep}^i(t))\}$. Here we only consider transfers if they are valid for the regular time $\tau_{reg}(e_{dep})$ of the departure event. Transfers that become valid due to delays are handled as part of the real-time reactions in Section~\ref{sec:Real-time reactions}. 

The expected perceived travel time for a selected boarding event $e_{dep}$ consists of the perceived travel time of the transfer $e_{arr}\rightarrow e_{dep}$ and $f_k(e_{dep})$. As we only consider valid transfers and the arrival time $\tau(e_{arr})$ is fixed at the moment of the decision, the probability $p_k^{delay}(e_{arr}\rightarrow e_{dep})$ of the transfer being invalid because of a delay is $0$. Let
\[f^{*}_k(e_{arr}) := \min_{e_{arr}\rightarrow e_{dep}, e_{dep}\in reldep(e_{arr})} ptt_k(e_{arr}\rightarrow e_{dep}) + f_k(e_{dep})\] be the perceived travel time for the optimal transfer among the relevant ones. The passenger first decides whether to walk to the destination or wait for a ride. The perceived travel time of the optimal transfer $f^{*}_k(e_{arr})$ is compared with the weighted walking time $walk_k^{dest}(s)$. 
Let $f_k^{opt}:=\min(walk_k^{dest}(s),f^{*}_k(e_{arr}))$ be the optimal decision. The probability that the passenger decides to walk to the destination is
\[p(walk):=\frac{e^{(f_k^{opt} - walk_k^{dest}(s))/\gamma(d)}}{e^{(f_k^{opt} - walk_k^{dest}(s))/\gamma(d)} + e^{(f_k^{opt} - f^{*}_k(e_{arr}))/\gamma(d)}},\]
If the passenger does not walk to the destination, he chooses a departure event from $reldep(e_{arr})$. We define the relative perceived travel time for a transfer $e_{arr}\rightarrow e_{dep}$ with $e_{dep}\in reldep(e_{arr})$ as 
\[f^{rel}_k(e_{arr}\rightarrow e_{dep}):=f^{*}_k(e_{arr})-(ptt_k(e_{arr}\rightarrow e_{dep}) + f_k(e_{dep})).\]
This value is therefore $0$ for the optimal transfer and negative otherwise. The probability that the passenger decides for the transfer $e_{arr}\rightarrow e_{dep}$ is
\[p(e_{arr}\rightarrow e_{dep}):=\frac{e^{f^{rel}_k(e_{arr}\rightarrow e_{dep})/\gamma(d)}}{\sum\limits_{e_{dep}\in reldep(e_{arr})} e^{f^{rel}_k(e_{arr}\rightarrow e_{dep})/\gamma(d)}}.\]
If a walk is required for the selected transfer, the passenger changes the stop after making the decision.
So far, we have assumed that the passenger has just alighted at an arrival event $e_{arr}$. However, it is also possible that a departure decision is made that does not immediately follow an alighting event. This is not a problem because any tuple $(s, \tau)$ consisting of a stop and a time can define a set of relevant boardings. We described the special case $(stop(e_{arr}),\tau(e_{arr}))$.

\paragraph*{Alighting decisions.}
If a passenger has boarded at a departure event $e_{dep}^i(t)$ of the trip $t$, he decides at which downstream stop of $t$ he will alight. He therefore decides on an arrival event $e_{arr}^j(t)$ with $j>i$. Together, $e_{dep}^i(t)$ and $e_{arr}^j(t)$ result in a trip segment $e_{dep}^i(t)\rightarrow e_{arr}^j(t)$ of the journey. The expected perceived travel time for a selected exit results from the sum of the perceived travel time of this trip segment and $f_k(e_{arr}^j(t))$. We again define the perceived travel time relative to the optimal decision. The perceived travel time for the optimal alighting decision is 
\[f^{*}_k(e_{dep}^i(t)) := \min_{m>i} ptt_k(e_{dep}^i(t)\rightarrow e_{arr}^m(t)) + f_k(e_{arr}^m(t)) .\] 
The relative perceived travel time for an exit at the $j$th stop is
\[f^{rel}_k(e_{dep}^i(t)\rightarrow e_{arr}^j(t)):=f^{*}_k(e_{dep}^i(t))-(ptt_k(e_{dep}^i(t)\rightarrow e_{arr}^j(t)) + f_k(e_{arr}^j(t))) .\]
This value is $0$ for the optimal decision and negative otherwise. The probability that the passenger chooses the $j$th stop is
\[p(e_{dep}^i(t)\rightarrow e_{arr}^j(t)):=\frac{e^{f^{rel}_k(e_{dep}^i(t)\rightarrow e_{arr}^j(t))/\gamma(d)}}{\sum\limits_{m>i} e^{f^{rel}_k(e_{dep}^i(t)\rightarrow e_{arr}^m(t))/\gamma(d)}}.\]

\paragraph{Choice-Set reduction.}
\label{sec:Choice-Set Reduction}
Not all possible events are relevant to passengers' decisions. We therefore reduce the choice set by filtering out unimportant events. The filtering is done before starting the simulation and is based on the regular timetable. This filtering has two motivations: on the one hand we want to prohibit unreasonable journeys, on the other hand we also want to improve the runtime and memory consumption of the simulation. The first rule is to discard journeys that are longer than the fastest journey in real time by an amount of $\Delta_{\tau}$. 

Trip segments $e_{dep}^i(t)\rightarrow e_{arr}^j(t)$ for which the minimum perceived travel time to the destination stop after alighting at $e_{arr}^j(t)$ is greater than before boarding at $e_{dep}^i(t)$ are not attractive for passengers. We want to filter out boardings at events $e_{dep}^i(t)$ for which every possible downstream exit from $t$ results in a trip segment that increases the minimum perceived travel time to the destination. Let $J^{opt}(e_{dep}^i(t))$ be the optimal journey that either starts by boarding $e_{dep}^i(t)$ or skips $e_{dep}^i(t)$ and waits at the current stop for another departure event. We filter out $e_{dep}^i(t)$ if $f_k(e_{arr}^{j}(t))>ptt(J^{opt}(e_{dep}^i(t)))$ for each $j>i$.

If $e_{dep}^i(t)$ is not discarded, we want to limit the possible alightings and subsequent footpaths. To keep the filtering ruleset independent of the exact boarding point of the current trip $t$, we define $J_j^{opt}(t):=\arg\max_{i<j}ptt(J^{opt}(e_{dep}^i(t))$ as the journey of maximum perceived travel time over journeys $J^{opt}(e_{dep}^i(t))$ at previous departures $e_{dep}^i(t)$. We filter out the possible exit at an arrival event $e_{arr}^{j}(t)$ if $f_k(e_{arr}^{j}(t))>ptt(J_j^{opt}(t))$. Since $J_j^{opt}(t)$ is the maximum over all previous departures, this inequality holds for all possible boarding points. Furthermore, we also want to limit the number of possible footpaths after an alighting event. Let $J^{opt}(e_{arr}^j(t), fp)$ with $fp=(stop(e_{arr}^j(t)),s')$ be the optimal journey that starts with alighting at $e_{arr}^j(t)$ and walking along the footpath $fp$. We ignore the footpath $fp$ after alighting at $e_{arr}^j(t)$ if $ptt(J^{opt}(e_{arr}^j(t), fp))>ptt(J_j^{opt}(t))$.

As a last rule, we want to filter out boardings at events $e_{dep}^i(t)$ that lead directly to loops in the journey. We discard $e_{dep}^i(t)$ if the optimal journey starting with $e_{dep}^i(t)$ includes the first departure event of $J^{opt}(e_{dep}^i(t))$ and $J^{opt}(e_{dep}^i(t))$ does not begin with boarding $e_{dep}^i(t)$. In this case the passenger would skip a loop by waiting for $J^{opt}(e_{dep}^i(t))$ instead of boarding $e_{dep}^i(t)$.

It is not sufficient to set the minimum perceived travel time $f_k$ of the discarded events to infinity, as events $e$ would still exist for which $f_k(e)<\infty$ applies, but which are no longer accessible to passengers. As these events would consume memory space, we remove all events that have become inaccessible. We apply the filter rules and discard inaccessible trips during a single additional CSA query.

\subsection{Learning}
\label{sec:Learning}

After each simulated day, the expected perceived travel times $f_k$ are updated according to the experiences on the current day. Thus, an explicit learning process is modeled through which the passengers learn properties of the network. We mark the learned properties with a tilde. The properties learned are the loads of driving arcs $\tilde{\lambda}_k(e_{dep}, e_{arr})$, the probabilities for denied boardings $\tilde{p}_k^{denied}(e_{dep})$, the probabilities for failed transfers because of delays $\tilde{p}_k^{delay}(e_{arr}\rightarrow e'_{dep})$ and the times $\tilde{\tau}_k(e)$ of the events. 
The learned values are updated in two steps: First, the expected properties are updated for the activities and events experienced by the passenger on the current day. Then the perceived travel times $f_k$ are recalculated. The recency parameter $\kappa$ indicates how highly new experiences are weighted. For $\kappa=1$, the accumulated experiences correspond to the average of all experiences. For $\kappa < 1$ newer experiences are weighted higher and for $\kappa > 1$ lower. The accumulated experiences are defined as in~\cite{cats2020} as a function of the experiences on the current day and the accumulated experiences before the current day. Since not every event is updated every day, we need to memorize the number of updates for each event and property. Let $|\lambda_k(e_{dep}, e_{arr})|, |p_k^{denied}(e_{dep})|$ and $|\tau_k(e)|$ be the number of updates. The respective number is incremented before the update.

At the end of each day, a passenger $k$ has a journey $\breakingcomma J_k=\{o\rightarrow e_{dep}^{i_1}(t_1), e_{dep}^{i_1}(t_1)\rightarrow e_{arr}^{j_1}(t_1),e_{arr}^{j_1}(t_1)\rightarrow e_{dep}^{i_2}(t_2),. ..., e_{dep}^{i_n}(t_n)\rightarrow e_{arr}^{j_n}(t_n),e_{arr}^{j_n}(t_n)\rightarrow dest\}$ and a set of denied boardings~$D_k$. Let $driving\_arcs(J_k)$ be the set of all driving arcs of $J_k$, $events(J_k)$ the set of all events of $J_k$ and $events_{dep}(J_k)$ the set of all departure events of $J_k$. In addition to the times of the events and the utilization of the activities, the proportion of passengers who were unable to board due to limited capacity out of those who attempted to on the current day $p_d^{denied}(e_{dep})$ is also stored for each departure event. 

Let $\lambda_d(e_{dep}, e_{arr}), p_d^{denied}(e_{dep})$ and $\tau_d(e)$ be the respective properties of the network on day $d$. 
The first step is to integrate the properties of the current day
into the learned values $\tilde{\lambda}_k(e_{dep}, e_{arr}), \tilde{p}_k^{denied}(e_{dep})$ and $\tilde{\tau}_k(e)$. 

We update the load $\tilde{\lambda}_k(e_{dep}, e_{arr})$ for the driving arcs $(e_{dep}, e_{arr})\in driving\_arcs(J_k)$:
\[\tilde{\lambda}_k(e_{dep}, e_{arr})\gets \tilde{\lambda}_k(e_{dep}, e_{arr})\cdot (1-|\lambda_k(e_{dep}, e_{arr})|^{-\kappa}) + \lambda_d(e_{dep}, e_{arr})\cdot |\lambda_k(e_{dep}, e_{arr})|^{-\kappa},\]
for the departure events $e_{dep}\in events(J_k)\cup F_k$ the probabilities $\tilde{p}_k^{denied}(e_{dep})$:
\[\tilde{p}_k^{denied}(e_{dep})\gets \tilde{p}_k^{denied}(e_{dep})\cdot (1-|p_k^{denied}(e_{dep})|^{-\kappa}) + p_d^{denied}(e_{dep})\cdot |p_k^{denied}(e_{dep})|^{-\kappa},\]
and for the events $e\in events(J_k)$ the times $\tilde{\tau}_k(e)$:
\[\tilde{\tau}_k(e)\gets \tilde{\tau}_k(e)\cdot (1-|\tau_k(e)|^{-\kappa}) + \tau_d(e)\cdot |\tau_k(e)|^{-\kappa}.\]

We set the learned load of a dwelling arc $(e_{arr}^i(t), e_{dep}^i(t))$ to the learned load of the following driving arc, i.e. $\tilde{\lambda}_k(e_{arr}^i(t), e_{dep}^i(t))\leftarrow \tilde{\lambda}_k(e_{dep}^i(t), e_{arr}^{i+1}(t))$.
In order to learn the probability $\tilde{p}_k^{delay}(e_{arr}\rightarrow e_{dep})$ that a transfer fails because $e_{arr}$ is delayed, the experienced times for each arrival event in $events(J_k)$ are memorized. Let $\mathcal{T}^d(e_{arr})$ be the sampled arrival times of $e_{arr}$ after day $d$. We calculate an inverse weighted empirical distribution function based on this sample. The probability $\tilde{p}_k^{delay}(e_{arr}\rightarrow e_{dep})$ is equal to the weighted share of sampled arrival times that are greater than $\tau_{reg}(e_{dep})-\tau_{min}(e_{arr}\rightarrow e_{dep})$, i.e. 
\[\tilde{p}_k^{delay}(e_{arr}\rightarrow e_{dep})=\sum_{i=1}^d w_i^d \cdot \mathds{1}_{\mathcal{T}_i>\tau_{reg}(e_{dep})-\tau_{min}(e_{arr}\rightarrow e_{dep})},\]
where $\tau_{min}(e_{arr}\rightarrow e_{dep})$ is the minimum required time for the transfer (either minimum change time or footpath length) and $\mathds{1}_{\mathcal{T}_i>\tau_{reg}(e_{dep})-\tau_{min}(e_{arr}\rightarrow e_{dep})}$ is the indicator for $\mathcal{T}_i>\tau_{reg}(e_{dep})-\tau_{min}(e_{arr}\rightarrow e_{dep})$ and
\[w_i^d = i^{-\kappa}\cdot \prod_{j=i+1}^d(1-j^{-\kappa}).\]
These normalized weights are derived from the update formulas above.
A problem arises when updating the times: as only parts of a trip are updated, it may happen that $\tilde{\tau}_k(e_1) > \tilde{\tau}_k(e_2)$ applies to an activity $(e_1, e_2)$. The property that the times of a trip are non-decreasing is therefore violated. To restore this property, the delays at the first or last event are extrapolated to the start or end of each trip.

After the values $\tilde{\lambda}_k, \tilde{p}_k^{denied}, \tilde{p}_k^{delay}$ and $\tilde{\tau}_k$ have been updated, the expected perceived travel times $f_k$ must be updated. However, a complete recalculation is not necessary as only a small subset of all events and activities have changed.

\subsection{Real-time Reactions}
\label{sec:Real-time reactions}
Until now, passengers' decisions have been based on personal experience or, in the absence of experience, on the timetable and default values. However, current circumstances may differ from these experiences. Therefore, mechanisms are implemented that allow adaptive behavior based on current information. Similar to Milatras~\cite{wahba2014}, these mechanisms allow to change the original decision. In general, a new decision is made if the currently selected action is worse than expected or if an unselected action is better than expected. The affected perceived travel times $f(e)$ are updated with a CSA query, taking into account the current information. The new decision then follows the same principle as the original, but with updated scores for each option. Boarding and alighting decisions are reconsidered for various cases, mainly due to differences between expected and actual event times or vehicle loads. 

We distinguish between real-time reactions while a passenger is waiting at a stop and real-time reactions while a passenger is in a vehicle. In general, a new decision is made if the currently selected action is worse than expected or if an unselected action is better than expected. The various cases are summarized in Table~\ref{tbl:realtimereactions}.

\begin{table}[t]
\centering
\caption{Routines which allow the passengers to change their original decision. The routines are executed at certain events if the conditions listed are met. The original decision is marked with a ${}^*$.}
\label{tbl:realtimereactions}
\begin{tabular}{llr}
\hline
Routine & Trigger-Event & Condition \\ \hline
boarding\_redo & $e_{dep}^{i}(t^*)$ & $\tau(e_{dep}^i(t^*))>\tilde{\tau}_k(e_{dep}^i(t^*))$ \\
boarding\_switch & $e_{dep}^j(t),~t\neq t^*$ & \makecell[tr]{$\tau(e_{dep}^j(t))<\tilde{\tau}_k(e_{dep}^j(t))$ \textbf{or} \\ $\tilde{p}_k^{denied}(e_{dep}^j(t))>0$ \textbf{or} \\ $\tau_{reg}(e_{dep}^j(t)) < \tau_{arr}$ \textbf{or} \\ $\tilde{\tau}_k(e_{dep}^{i}(t^*)) < \tau_{curr}$ } \\
alighting\_redo & $e_{arr}^{i^*}(t)$ & $(e_{arr}^{i^*}(t)\rightarrow e_{dep}^{opt})$ is invalid\\
alighting\_switch & $e_{arr}^{i}(t),~i<i^*$ & \makecell[tr]{$\tau(e_{arr}^i(t))<\tilde{\tau}_k(e_{arr}^i(t))$ \textbf{or} \\ $(e_{arr}^{i^*}(t)\rightarrow e_{dep}^{opt})$ is invalid \textbf{or} \\ passenger is unexpectedly standing}\\
\hline
\end{tabular}
\end{table}

For real-time reactions at stops, we further distinguish between two cases. In the first case, let the originally selected trip $t^*$ be the currently departing trip (boarding\_redo). If $\tau(e_{dep}^i(t^*))>\tilde{\tau}_k(e_{dep}^i(t^*))$ applies, the trip is more delayed than expected and a change of decision is possible. After updating $f_k(e_{dep}^i(t^*))$, a completely new boarding decision is made as described in Section~\ref{sec:choicemodel}. The current trip $t^*$ remains a possible option. 

For the second case, let $e_{dep}^j(t)\neq e_{dep}^{i}(t^*)$. The passenger therefore has the option of switching from the currently selected trip $t^*$ to trip $t$ (boarding\_switch). We distinguish between four different reasons why a switch could be advantageous:
\begin{enumerate}
	\item trip $t$ is departing earlier than expected,
	\item there is still free capacity in $t$ and $\tilde{p}_k^{denied}(e_{dep}^j(t))>0$ applies,
	\item boarding at $e_{dep}^j(t)$ is not possible according to regular times and was therefore not included in the original decision, and 
	\item the current simulation time $\tau_{curr}$ is greater than the expected departure time of the current choice~$t^*$, i.e., the current choice is more delayed than the passenger expected.
\end{enumerate}

A binary decision between these two options is made following the described $(\epsilon-greedy, SoftMax)$ decision model with two restrictions after updating the values $f(e_{dep}^j(t))$ and $f(e_{dep}^i(t^*))$. To avoid a bias towards earlier departing but worse trips, the switch is only executed if the updated perceived travel time for $e_{dep}^j(t)$ is not worse than the updated perceived travel time for $e_{dep}^i(t^*)$. On the other hand, we also want to avoid that the passenger switches to a better trip if the original choice was already suboptimal, as this would defeat the purpose of a probabilistic decision model. The switch is therefore not made if the original choice was suboptimal and the original choice is not worse than expected under the current circumstances.

The real-time reactions in vehicles work in a similar way to the real-time reactions at stops. The current passenger $k$ has a currently selected arrival event $e_{arr}^{i^*}(t)$ at which they want to alight. If the trip $t$ arrives at a stop, the passenger has the option to change his decision. Let $e_{arr}^{i}(t)$ with $i\leq i^*$ be the current arrival event. 
In the case where $i=i^*$, the passenger has arrived at the stop where he plans to alight. A change of decision is possible if the current trip is more delayed than expected. If the optimal transfer is still possible, no change of decision is necessary, as the journey with the minimum perceived travel time to the destination remains the same. If this transfer is no longer possible, the perceived travel time for an exit at $e_{arr}^{i}(t)$ becomes worse. In this case, a new alighting decision is made, taking into account the current information (alighting\_redo). The current arrival remains a possible option.

If $i<i^*$ applies, the passenger has the option of alighting early at the current stop (alighting\_switch). There are three reasons why early alighting might be attractive: 
\begin{enumerate}
	\item the current trip $t$ arrived earlier than expected, which could allow a new transfer at the current stop,
	\item the current trip $t$ arrived later than expected and the optimal transfer at the originally selected exit at $e_{arr}^{i^*}(t)$ is no longer possible with the projected downstream delay, and 
	\item the passenger has no seat and $seated_k(e_{dep}^h(t),(e_{dep}^{i-1}(t), e_{arr}^i(t))$ was $true$ at the time of the original decision ($e_{dep}^h(t)$ being the boarding event), in which case the passenger expects to have to stand for the rest of the trip.
\end{enumerate}

If one of these causes is given, a binary decision is made between the two events $e_{arr}^{i}(t)$ and $e_{arr}^{i^*}(t)$. As with the decision at stops, the switch is only made if the expected perceived travel time for alighting at the current event $e_{arr}^{i}(t)$ is not worse than for the current choice $e_{arr}^{i^*}(t)$. If the original choice $e_{arr}^{i^*}(t)$ was suboptimal and the optimal transfer is still possible after the exit at event $e_{arr}^{i^*}(t)$ according to the current information, the switch is also not made. 
It is still possible for a passenger to find a seat sooner than they expected when they made their original decision. 
 In this case, later exits can become more attractive because the perceived travel time in the current trip is smaller than expected. If the passenger finds a seat, a new alighting decision is made.

%---------------------------------------------------
\section{Experimental Study} \label{sec:experiments}
In this section, the model is tested on two different public transportation networks. First, the experimental setup and the choice of parameters are presented. Subsequently, three different experiments are conducted 
to assess the proposed model.

\subsection{Experimental Setup}
The simulation was implemented in C++ and compiled with mvc 14.3 on Windows 10 using the O2 compiler option. The experiments were run on an AMD Ryzen 7 5800X, clocked at 4.7 GHz during program execution, with 32 GB DDR4 memory with a latency of CL16 and a clock frequency of 3600 MHz. As the model is probabilistic, the results are averaged over ten runs. We simulate a total of two hours of the timetable per day and evaluate all passengers who start their journey within the first simulated hour. For each run, we simulate $30$ days. 
The 2-hour timeframe is sufficient to demonstrate the learning process as the timetable is hourly periodic. As we only simulate a limited timeframe, it is not guaranteed that passengers can finish their journey, for example in the case of consecutive denied boardings. For this case, we introduce a new penalty $\beta^{unfinished}$, which we set to the Euclidean distance in meters between the current stop and the destination. The parameters for the perceived travel times are identical for each passenger $k$ and are chosen as $\beta_k^{wait}=1$, $\beta_k^{walk}=1.5$ and $\beta_k^{transfer}=300$. We choose $\beta_k^{fail}=2$ as the multiplier for the additional waiting or walking time after a failed boarding. The choice of the crowding factor $\beta_k^{crowding}$ is based on a British meta-study~\cite{wardman2011} and is summarized in Table~\ref{tbl:crowdingpenalty}. 

\begin{table}[bt]
\centering
\caption{Crowding factor $\beta_{k}^{crowding}$}
\label{tbl:crowdingpenalty}
\begin{tabular}{lrr}
\hline
load & seated & standing \\ \hline
$0\leq \lambda \leq 0.6$ & $1.0$ & \\
$0.6<\lambda \leq 1.0$ & $1.2$ & \\
$1.0<\lambda \leq 2.0$ & $1.4$ & $2.2$ \\
\hline
\end{tabular}
\end{table}

We use datasets for the public transportation networks of Göttingen (goevb) and Stuttgart~\cite{PTNOpenSource}, provided in LinTim format~\cite{SchieweSchoebelJaegeretal.2023}, including the OD matrix. Stuttgart is a mixed network, consisting of $25$ train lines and $378$ bus lines. Göttingen, in contrast, is a pure bus network consisting of $22$ lines. Table~\ref{tbl:instances} shows the main properties of the two data sets. 
The buses in Stuttgart have a total capacity of $70$ and the total capacity of the trains is between $400$ and $1000$. In Göttingen, buses have a capacity of $50$. For simplicity, we assume that the number of seats corresponds to half the total capacity. This is in line with common bus models. We assume that for buses $0.4$ passengers can board or alight per second ($doorCapacity$), corresponding to the value recommended by the Transportation Research Board for buses with two doors~\cite{nap}. 
For trains, let $doorCapacity=cap/200$. The value is chosen depending on the capacity,     based on a study for the French city of Nantes~\cite{christoforou2020}. The minimum transfer times $mct$ are specified by LinTim for both data sets. For Stuttgart, the minimum transfer time is $60 s$ and for Göttingen $180 s$. Footpaths were limited to a maximum of $1800 s$. We discard journeys that arrive more than $\Delta_{\tau}=3600 s$ later than the fastest journey on the current day.
The rest of the parameters have been determined by testing. We have chosen a constant temperature of $\gamma=400$ for Göttingen and $\gamma=250$ for Stuttgart. For both networks, we chose $\epsilon=0.2$ for the probability of choosing a random option. Furthermore, we have set the recency parameter to $\kappa=0.5$ and the standard load to $\lambda_{std}=0.5$.

\begin{table}[t]
\centering
%\captionsetup{format=plain}
\caption{Characteristics of the Stuttgart and Göttingen networks. The number of trips and driving arcs refer to the 2-hour simulation frame.}
\label{tbl:instances}
\begin{tabular}{lrrrrrr}
\hline
Network & \#stops & \#lines & \#trips & \#driving arcs & \#footpaths & \# passengers per hour\\ \hline
Göttingen & $257$ & $22$ & $205$ & $2348$ & $0$ & $1943$\\
Stuttgart & $735$ & $403$ & $3175$ & $17381$ & $8732$ & $44836$\\
\hline
\end{tabular}
\end{table}

\subsection{Performance}
The simulation was run in parallel on $16$ threads. For Stuttgart, an average runtime of $493$ seconds was achieved for the whole simulation. Calculating the initial perceived travel times took $70$ seconds. On a single day, $89672$ passengers were simulated in less than $14.1$ seconds on average. For Göttingen, the whole simulation took about $6$ seconds. The memory consumption of the simulation is relatively high, as each passenger has to store his or her personal experiences and expected travel times. For Stuttgart $10.5$~GB and for Göttingen $0.2$ GB were used.

\subsection{Experiments}

\begin{figure}
\begin{subfigure}{.49\textwidth}
  \centering
  \includegraphics[width=.9\linewidth]{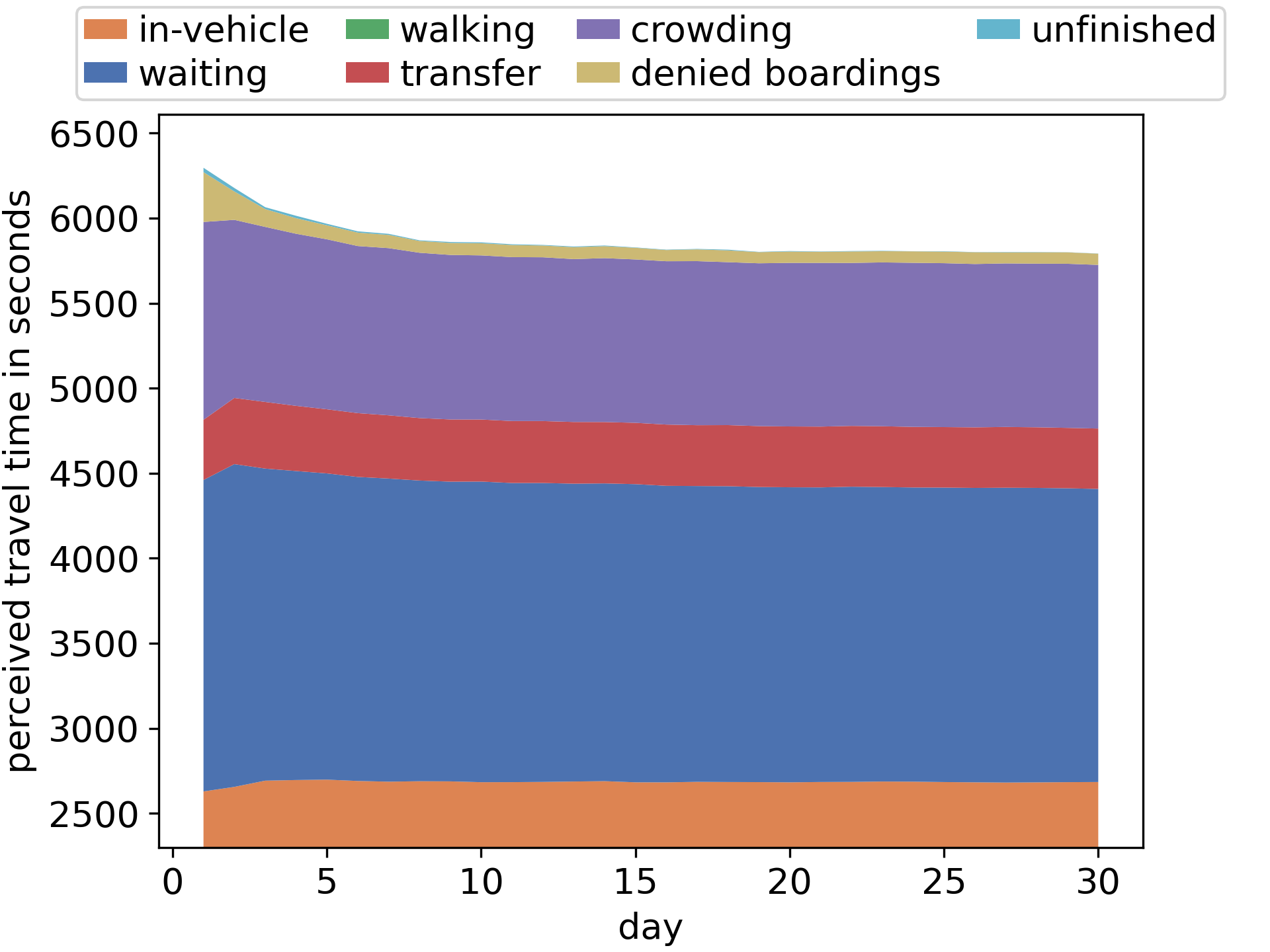}
  \caption{Göttingen}
  \label{fig:exp1_goevb_route_costs}
\end{subfigure} \hfill
\begin{subfigure}{.49\textwidth}
  \centering
  \includegraphics[width=.9\linewidth]{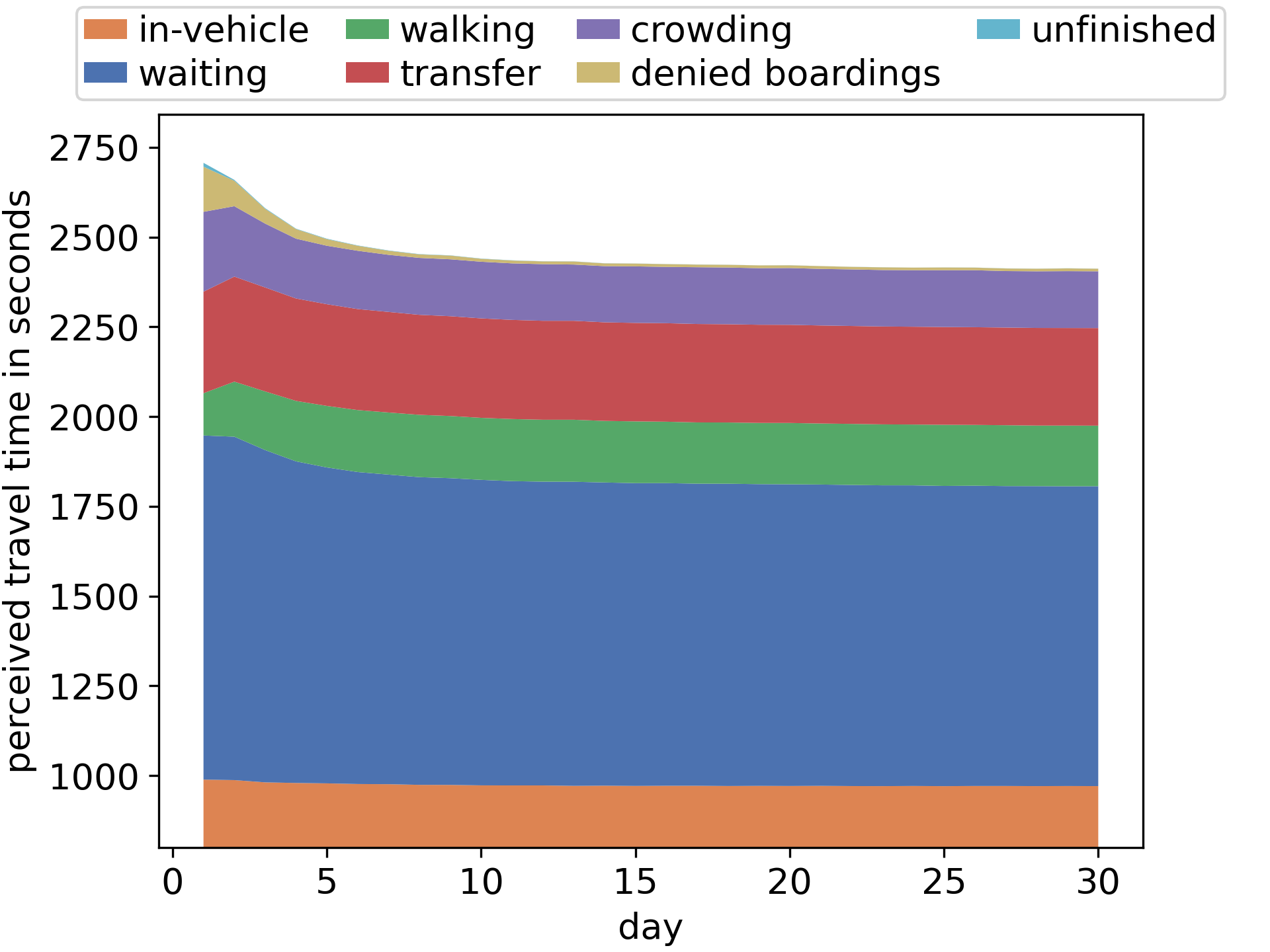}
  \caption{Stuttgart}
  \label{fig:exp1_suttgart_route_costs}
\end{subfigure}
\caption{Evolution and composition of the perceived travel times.}
\label{fig:exp1_route_costs}
\end{figure}

\paragraph{Experiment 1: Evolution of perceived travel times.}
In a first experiment, we examine the evolution of the average perceived travel times over 30~days. 
For simplicity, we assume that all passengers start their learning process on the first day. More complex scenarios are possible in the model. 
This evolution is shown for both networks in Figure~\ref{fig:exp1_route_costs}, including the components of the perceived travel times. Table~\ref{tbl:penalty_details} provides the detailed composition of the perceived travel times for the first and last day of the experiment. 

The overall average perceived travel time decreases from $6296s$ to $5792s$ for Göttingen and from $2707s$ to $2412s$ for Stuttgart. Perceived travel times decrease significantly in the first few days. After $10$ days, only minimal gains are achieved. The variance between each of the ten runs was quite small with a maximum deviation from the mean improvement of about $10\%$ for Göttingen and $6\%$ for Stuttgart.

The three effects of network congestion are directly reflected in the additional waiting time for denied boardings, the real waiting time and the penalty term for overcrowded vehicles. For both networks, these components of the total perceived travel time decrease over the course of the simulation. The greatest difference is recorded in the additional waiting time penalty for denied boardings. For Göttingen, this value decreases by $227s$ and for Stuttgart by $118s$. The average number of denied boardings per passenger decreases from $0.24$ to $0.05$ for Göttingen and from $0.21$ to $0.01$ for Stuttgart.

For both networks, the real waiting time decreases significantly during the simulation ($108s$ decrease for Göttingen and $123s$ decrease for Stuttgart). For Göttingen, it increases on the first few days and only drops below the initial value on the fourth day. The reason for this is that passengers explore nominally worse journeys due to the experienced network congestion effects. Similar effects can be seen in the number of transfers, walking time and the real travel time. If the nominally worse journeys are similarly congested as the journeys selected on previous days, the overall perceived travel time can increase.
The crowding penalty decreases by $201s$ for Göttingen and $64s$ for Stuttgart. Most of this improvement is due to passengers' desire to avoid standing. The average standing time drops from $495s$ to $307s$ for Göttingen and from $69s$ to $31s$ for Stuttgart.
In this experiment, we have shown how passengers respond to network congestion through the learning process and improve their average perceived travel time by incorporating personal experiences and avoiding congestion. We have found that denied boardings and the resulting additional waiting times have the largest impact on passengers. 

\begin{table}
\centering
\caption{Detailed composition of the perceived travel times for the first and last day (in seconds).}
\label{tbl:penalty_details}
\begin{tabular}{lrrrrrrrr}
\hline
Network & total & in-vehicle & wait & walk & transfer & crowding & denied & unfinished\\
        &       &            &      &      &          & & boarding &\\
 \hline
Göttingen day $1$ & $6296$ & $2628$ & $1832$ & $0$ & $354$ & $1163$ & $293$ & $26$\\
Göttingen day $30$ & $5792$ & $2684$ & $1724$ & $0$ & $355$ & $962$ & $66$ & $1$\\
\addlinespace[0.2cm]
Stuttgart day $1$ & $2707$ & $989$ & $959$ & $78$ & $283$ & $222$ & $126$ & $11$\\
Stuttgart day $30$ & $2412$ & $970$ & $835$ & $113$ & $272$ & $158$ & $7$ & $0$\\
\hline
\end{tabular}
\end{table}

\begin{figure}
\begin{subfigure}{.49\textwidth}
  \centering
  \includegraphics[width=.9\linewidth]{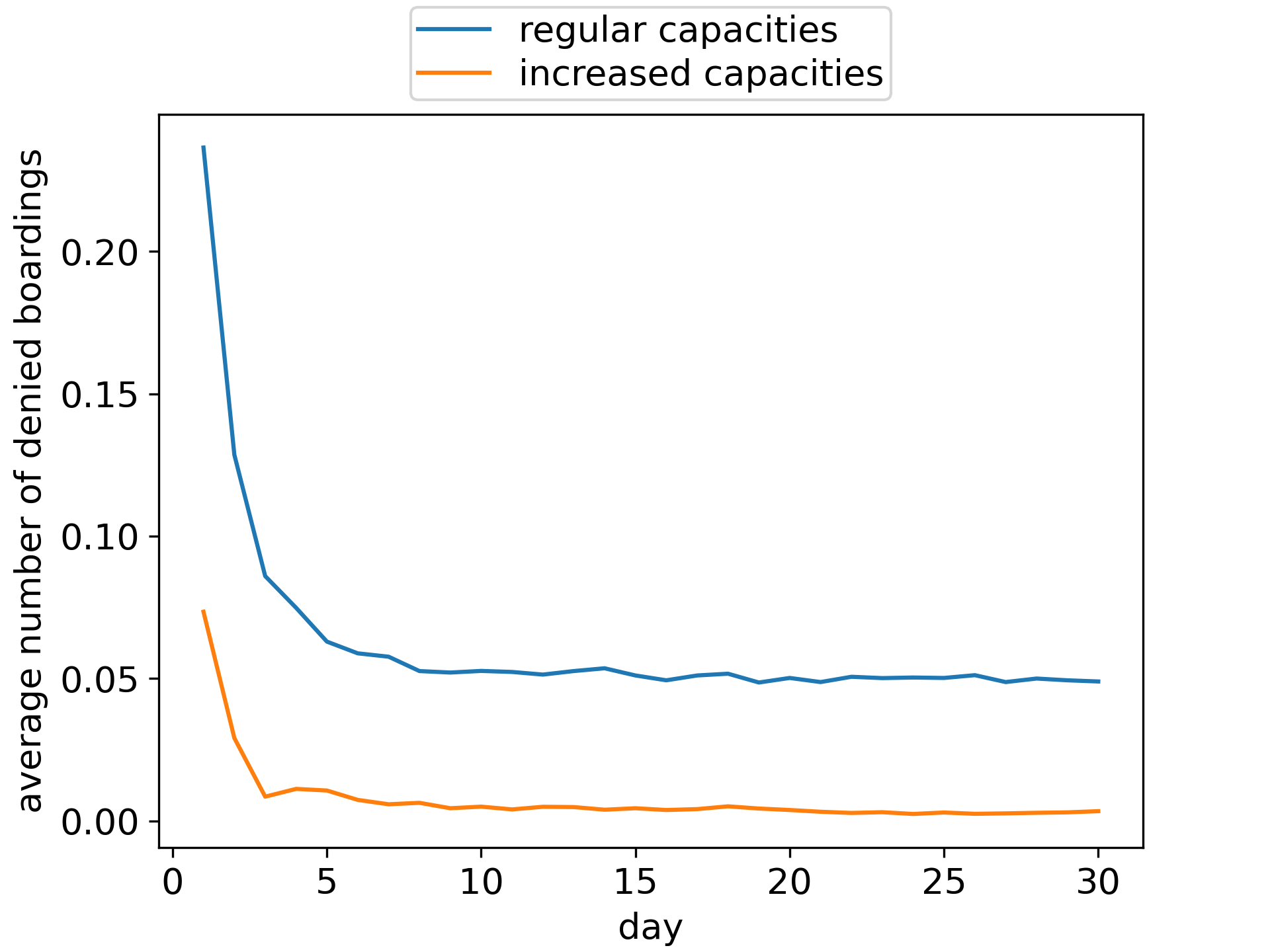}
  \caption{Göttingen}
  \label{fig:exp2_goevb_fail_to_board_events}
\end{subfigure}\hfill
\begin{subfigure}{.49\textwidth}
  \centering
  \includegraphics[width=.9\linewidth]{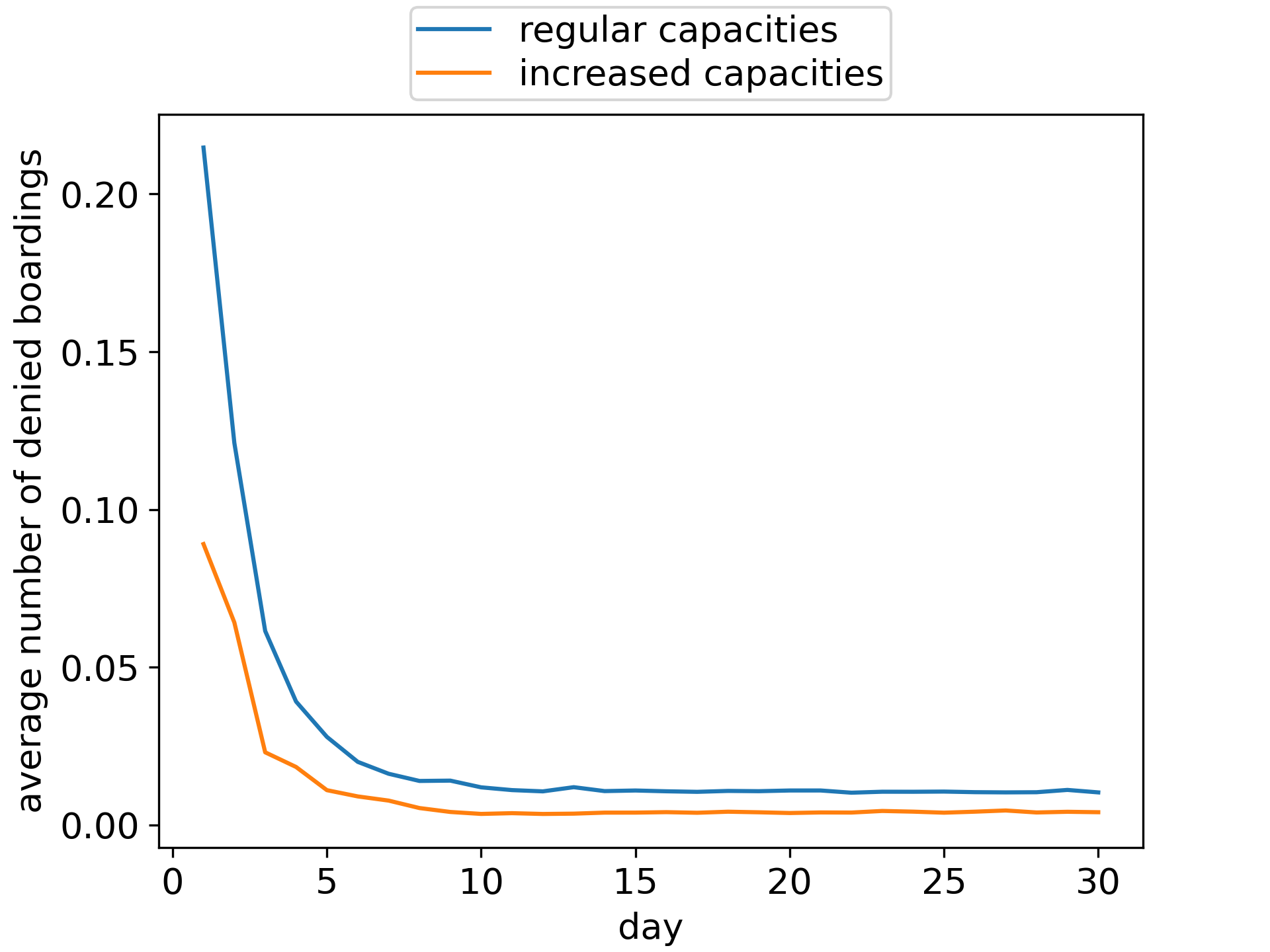}
  \caption{Stuttgart}
  \label{fig:exp2_suttgart_fail_to_board_events}
\end{subfigure}
\caption{Evolution of denied boardings per passenger with regular and increased capacities.}
\label{fig:exp2_fail_to_board_events}
\end{figure}

\paragraph{Experiment 2: Capacity expansion.}
In this experiment, we examine the benefits of increasing capacity, targeted to trips operating on full capacity. 
To determine candidates for a capacity increase, we simulate both networks for one day and determine the trips that are affected by denied boardings. Each trip with at least one denied boarding has its capacity increased by $40\%$. As a result, the capacities were increased for $41$ trips in Göttingen and for $236$ trips in Stuttgart. 
A large benefit was achieved by this capacity expansion for both networks, especially on the first day of the simulation. Compared to the first day of Experiment 1, the total perceived travel time for Göttingen decreased from $6296s$ to $5775s$. For Stuttgart, the difference is smaller (from $2707s$ to $2585s$). As the simulation progresses, this difference becomes smaller as passengers adjust their behavior when capacity is at its limit. At the end of the simulation, the difference compared to Experiment 1 is $321$ seconds for Göttingen and $31$ seconds for Stuttgart.
The improvement is largely due to the reduction in the number of denied boardings and the resulting improvement in real waiting times. The average number of denied boardings is shown in Figure~\ref{fig:exp2_fail_to_board_events}. On the final day, the average number of denied boardings is below $0.01$ for both networks, which is a substantial improvement for Göttingen. This suggests that for Göttingen the regular capacity is too limited to satisfy passenger demand.

\begin{figure}
\begin{subfigure}{.45\textwidth}
  \centering
  \includegraphics[width=.85\linewidth]{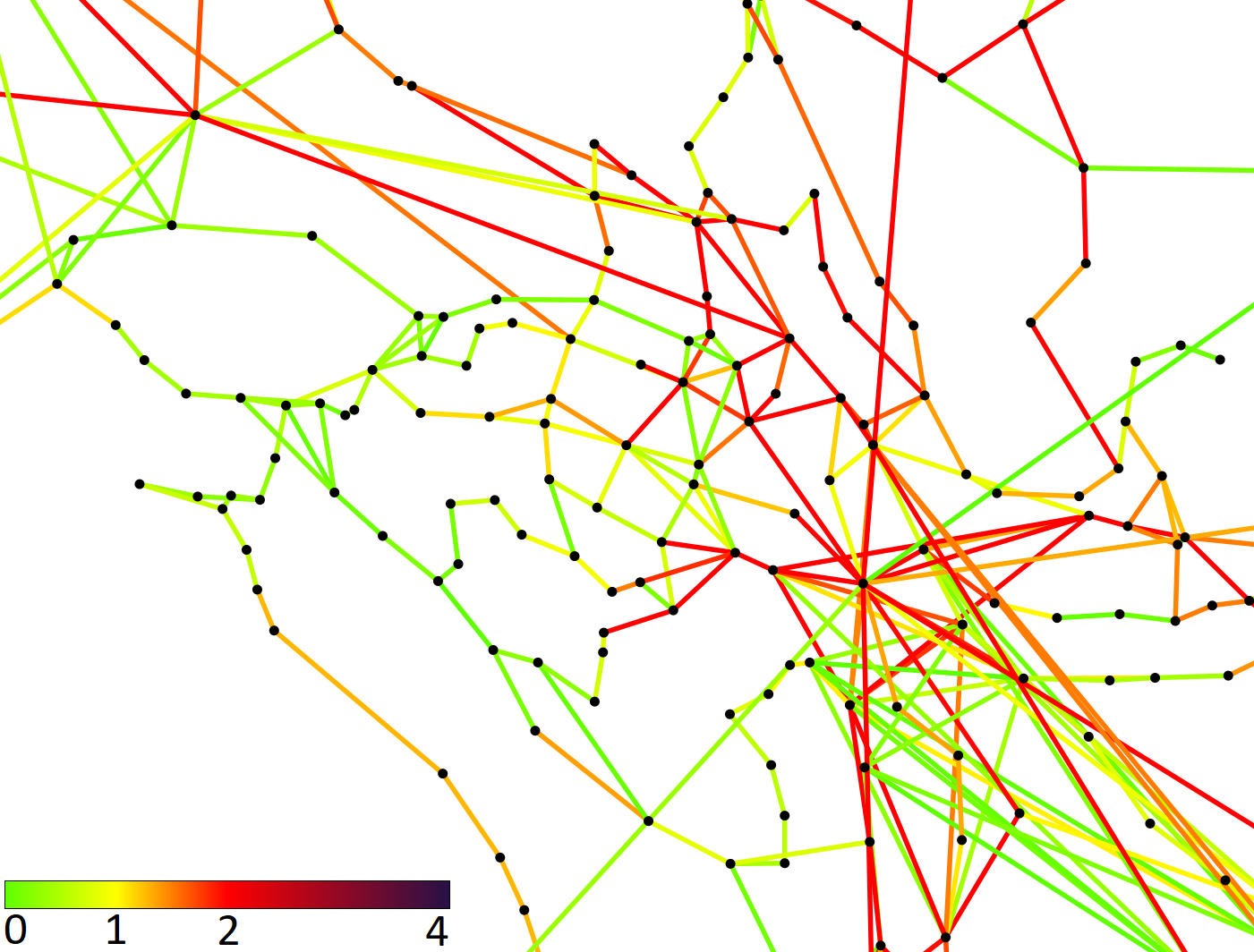}
  \caption{normal vehicle capacities}
  \label{fig:exp3_loads_a}
\end{subfigure}\hfill
\begin{subfigure}{.45\textwidth}
  \centering
  \includegraphics[width=.85\linewidth]{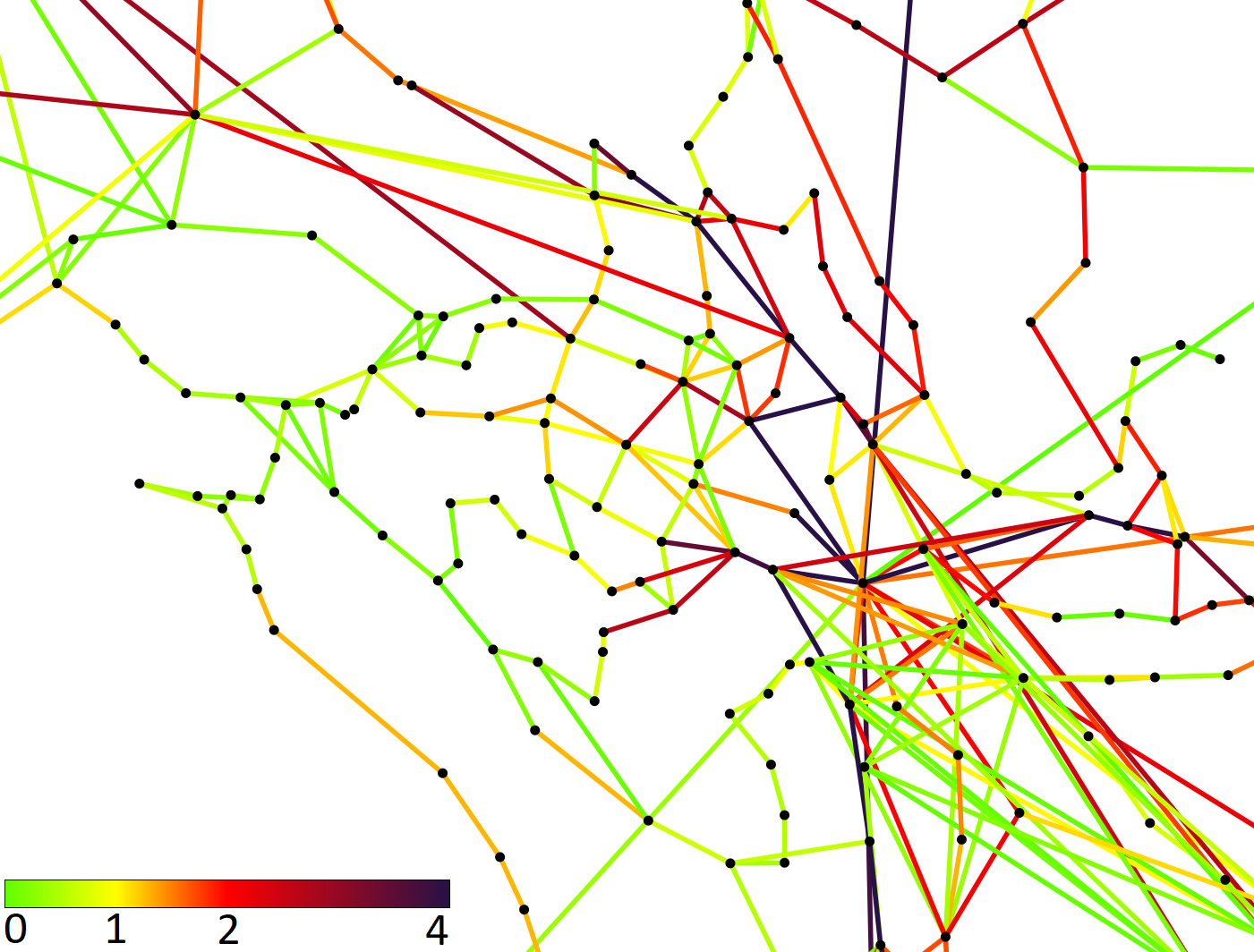}
  \caption{unlimited vehicle capacities}
  \label{fig:exp3_loads_b}
\end{subfigure}
\caption{Small excerpt from the Stuttgart network. Comparison of vehicle loads: normal (left) vs.\ unlimited vehicle capacities (right). 
The maximum capacity utilization on the respective segment is shown.
The color coding is relative to the seat capacity. The value 2.0 corresponds to full capacity and 1.0 to full seat capacity.}
\label{fig:exp3_loads}
\end{figure}

\paragraph{Experiment 3: Unlimited capacities.}
In Experiment 3, we study the scenario of unlimited vehicle capacities.
As a result, some trips are highly overloaded. For Stuttgart, there are buses with over $300$ passengers and trains with over $1000$ passengers. Similarly, buses with over $130$ passengers are found in Göttingen. For Göttingen, $7.9\%$ of all driving edges have loads greater than their regular capacity, compared to $2.6\%$ for Stuttgart.
In Figure~\ref{fig:exp3_loads}, we compare the vehicle load of normal capacities (left) and unlimited capacities (right). 
The color coding is relative to the seat capacity. The value 2.0 corresponds to full capacity and 1.0 to full seat capacity. Values above 2.0 indicate overload. We observe that similar segments of the network have high loads in both cases, but considering capacities avoids overloading.   

\begin{figure}[tb]
\centering
\includegraphics[width=.8\linewidth]{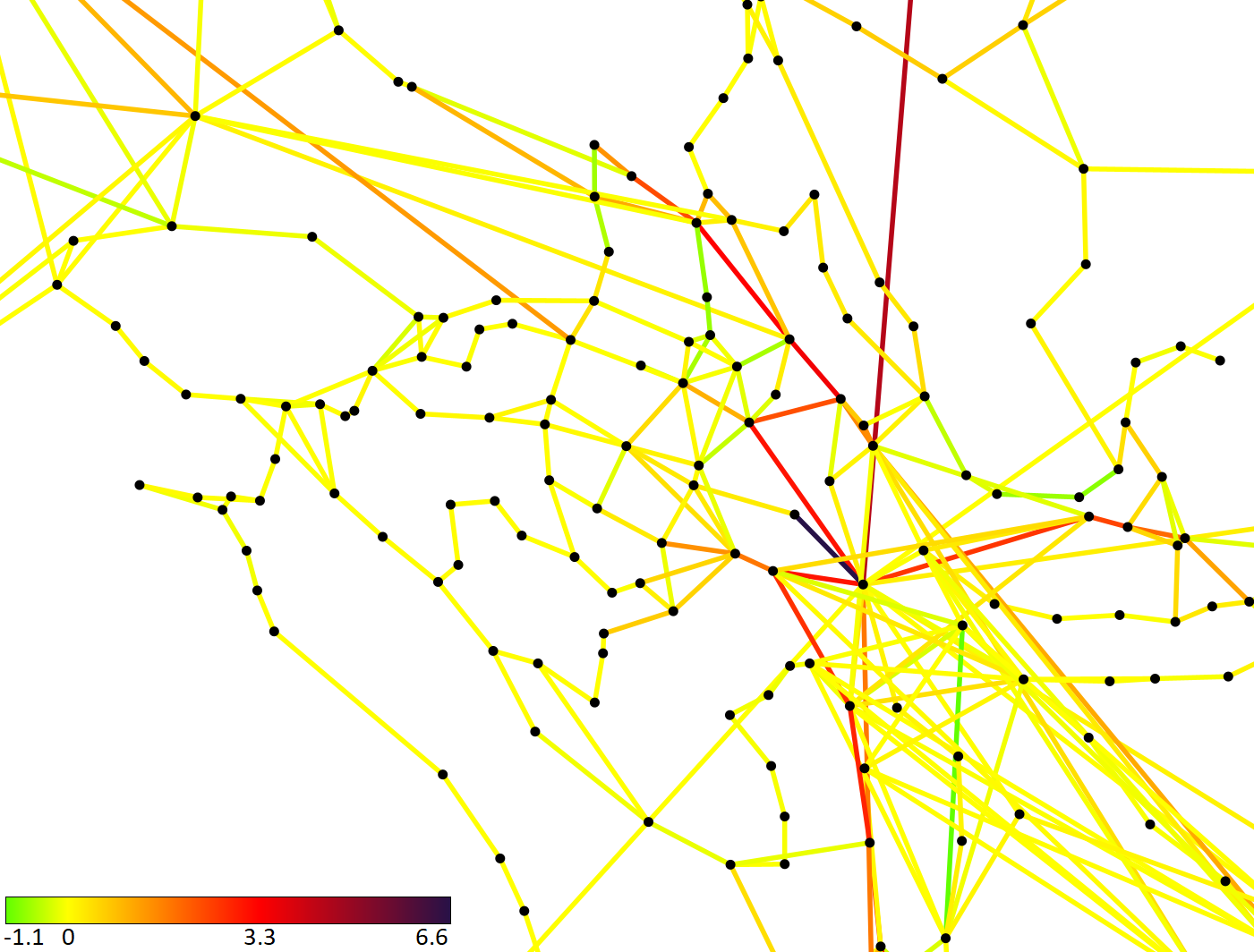}
\caption{\label{fig:exp3_diff}Excerpt from the Stuttgart network. We show the difference of loads of the scenario with unlimited and with limited capacities. Red shades mean that load has been reduced; yellow shades mean that it remains essentially unchanged, green shades mean that it has increased.
}
\end{figure}

To make the differences of the resulting traffic assignments clearer, in Figure~\ref{fig:exp3_diff} we highlight the difference of loads between unlimited and limited capacity. Red shades mean that load has been reduced; yellow shades mean that it has remained essentially unchanged and green shades mean that it has increased. We can therefore see that the overload on certain segments in the unlimited capacity scenario has been shifted to other segments. However, and not visible in this figure, most of the shift in load is from highly crowded trips to less crowded trips on the same line. 

\vspace*{-2ex}
%---------------------------------------------------
\section{Conclusions and Outlook}\label{sec:outlook}

We presented a fine-grained framework for a dynamic agent-based simulation of traffic assignment in public transit networks. 
This model is extendible to include real-time delay information or real-time load rates. First experimental studies with our prototype prove to be highly efficient for simulation tests with medium-sized metropolitan regions. 
As part of future work, further case studies are needed to assess the model validity and scalability to even larger networks as well as to calibrate model parameters.

%---------------------------------------------------

\bibliography{literature}

\end{document}